\documentclass{elsart}

	\usepackage{graphicx}
	\usepackage{amsmath}
	\usepackage{amsfonts}
	\usepackage{amssymb}
	\usepackage{latexsym}
	\usepackage{mathrsfs}
	\usepackage{epstopdf}
	\usepackage{tabularx}

	\newtheorem{assumption}{Assumption}
	\newtheorem{lemma}{Lemma}
	\newtheorem{proposition}{Proposition}
	
	\newtheorem{corollary}{Corollary}

	{
	\begin{bmatrix}}%
	{\end{bmatrix}
	}

	 \def\build#1_#2^#3{\mathrel{\mathop{\kern0pt#1}\limits_{#2}^{#3}}}%

\begin{document}

\begin{frontmatter}

\title{An MPC approach to output-feedback control of stochastic linear discrete-time systems} % Title, preferably not more than 10 words.

\author[poli]{Marcello Farina,}
\author[poli]{Luca Giulioni,}
\author[magni]{Lalo Magni,}
\author[poli]{and Riccardo Scattolini}

\address[poli]{Dipartimento di Elettronica, Informazione e Bioingegneria, Politecnico di Milano, Milan, Italy: \{marcello.farina,luca.giulioni,riccardo.scattolini\}@polimi.it}
\address[magni]{Dipartimento di Ingegneria Civile e Architettura, Universit\`a di Pavia, Pavia, Italy: lalo.magni@unipv.it}

% \begin{keyword}                           % Five to ten keywords,
% stochastic control, predictive control,              % chosen from the IFAC
% \end{keyword}                             % keyword list or with the
                                          % help of the Automatica
                                          % keyword wizard

\begin{abstract}                          % Abstract of not more than 250 words.
In this paper we propose an output-feedback Model Predictive Control (MPC) algorithm for linear discrete-time systems affected by a possibly unbounded additive noise and subject to probabilistic constraints. In case the noise distribution is unknown, the chance constraints on the input and state variables are reformulated by means of the Chebyshev - Cantelli inequality. The recursive feasibility of the proposed algorithm is guaranteed and the convergence of the state to a suitable neighbor of the origin is proved under mild assumptions. The implementation issues are thoroughly addressed showing that, with a proper choice of the design parameters, its computational load can be made similar to the one of a standard stabilizing MPC algorithm. Two examples are discussed in details, with the aim of providing an insight on the performance achievable by the proposed control scheme.
\end{abstract}
\end{frontmatter}
\section{Introduction}
The problem of designing robust deterministic Model Predictive Control (MPC) schemes, has nowadays many solutions, see for example \cite{MagniBook07,rawbook,LimonEtAl-NMPC09}. However, the proposed approaches are in general computationally very demanding, since they either require the solution to difficult on-line min-max optimization problems, e.g.,~\cite{MagniDeNScatAll03}) or the off-line computations of polytopic robust positive invariant sets, see \cite{mayne2005robust}. In addition they are conservative, mainly because they (implicitly or explicitly) rely on worst-case approaches. Moreover, in case the uncertainties/disturbances are characterized as stochastic processes, constraints must be reformulated in a probabilistic framework \cite{Yaesh2003351,Hu2012477}, worst-case deterministic methods do not take advantage of the available knowledge on the characteristics of the process noises, such as their probability density function, and cannot even guarantee recursive feasibility in case of possibly unbounded disturbances.\\
Starting from the pioneering works \cite{Schwarm99,Li02}, these reasons have motivated the development of MPC algorithms for systems affected by stochastic noise and subject to probabilistic state and/or input constraints. Mainly two classes of algorithms have been developed so far. The first one relies on the randomized, or scenario-based approach, see e.g.,  \cite{Batina,Calaf,BlackmoreOnoBektassovWilliams-TR10}, a very general methodology that allows to consider linear or nonlinear systems affected by noise with general distributions with possibly unbounded and non-convex support. As a main drawback, randomized methods are still computationally very demanding for practical implementations and their feasibility and convergence properties are difficult to prove.\\
The second approach, referred in \cite{Cogill} as probabilistic approximation method, is based on the point-wise reformulation of probabilistic, or expectation, constraints in deterministic terms to be included in the MPC formulation. Interesting intermediate methods have been proposed in \cite{BernardiniBemporad-TAC12}, where a finite number of disturbance realizations are assumed, and in \cite{Korda}, where constraints averaged on time are considered. Among the wide class of probabilistic approximation algorithms, a further distinction can be based on the noise support assumptions, which can be either bounded, e.g., as in \cite{Korda,KouvaritakisCannonRakovicCheng-Automatica10,CannonChengKouvaritakisRakovic-Automatica12}
or unbounded, see for instance \cite{Bitmead,Primbs,Hokaymen,Cogill,CannonKouvaritakisWu-Automatica09,Ono-ACC12}. While for bounded disturbances recursive feasibility and convergence can be established, the more general case of unbounded noise poses more difficulties and some specific solutions and reformulations of these properties have been adopted, for example in \cite{CannonKouvaritakisWu-Automatica09} the concept of invariance with probability $p$ is used, while in \cite{Ono-ACC12} the definition of probabilistic resolvability is introduced. Also, linear systems with known state have generally been considered, with the notable exceptions of \cite{Bitmead,Hokaymen,CannonChengKouvaritakisRakovic-Automatica12}, where output feedback methods have been proposed.\\
Finally, it must be remarked that some of the mentioned approaches have been successfully applied in many applicative settings, such as building temperature regulation \cite{OldewurtelParisioJonesMorariEtAl-ACC10} and automotive applications \cite{GrayBorrelli-ITSC13,BlackmoreOnoBektassovWilliams-TR10,BichiRipaccioliDiCairanoBernardiniBemporadKolmanovsky-CDC10}.\\
In this paper, an output feedback algorithm for linear discrete-time systems affected by a possibly unbounded additive noise is proposed. In case the noise distribution is unknown, the chance constraints on the inputs and state variables are reformulated by means of the Chebyshev - Cantelli inequality \cite{Cantelli}, as originally proposed in \cite{Locatelli} for the design of decentralized controllers and in \cite{Pala} in the context of MPC. Later, this approach has also been considered in \cite{Cogill,GrayBorrelli-ITSC13}, and used to develop preliminary versions of the algorithm here proposed in \cite{FGMS13_CDC,FGMS14_IFAC}. With respect to \cite{FGMS13_CDC,FGMS14_IFAC}, in this paper we discuss, in a consistent fashion and in a detailed way, our control approach. In particular, we address also the case when the noise distribution is known (i.e., and it is Gaussian). We also address algorithm implementation aspects, proposing two novel and theoretically well funded approximated schemes and full implementation details. The algorithm computational load can be made similar to the one of a standard stabilizing MPC algorithm with a proper choice of the design parameters. Importantly, the computation of robust positively invariant sets is not required and, in view of its simplicity and of the required lightweight computational load, the application of the proposed approach to medium/large-scale problems is allowed. The recursive feasibility of the proposed algorithm is guaranteed by a switching MPC strategy which does not require any relaxation technique, and the convergence of the state to a suitable neighbor of the origin is proved.\\
The paper is organized as follows. In Section \ref{sec:problem_statement} we first introduce the main control problem, then we define and properly reformulate the probabilistic constraints.
In Section \ref{MPC} we formulate the Stochastic MPC optimization problem and we give the general theoretical results. Section \ref{sec:num_implementation} is devoted to the implementation issues, while in Section \ref{sec:example} two examples are discussed in detail: the first one is analytic and is aimed at comparing the conservativeness of the algorithm to the one of the well known tube based approach \cite{mayne2005robust}, while the second one is numeric and allows for a comparison of the different algorithm implementations. Finally, in Section \ref{sec:conclusions} we draw some conclusions. For clarity of exposition, the proof of the main theoretical results is postponed to the Appendix.\\
\textbf{Notation}. The symbols $\succ$ and $\succeq$ (respectively $\prec$, and $\preceq$) are used to denote positive definite and semi-positive definite (respectively negative definite and semi-negative definite) matrices. The point-to-set distance from $\zeta$ to $\mathcal{Z}$ is $\mathrm{dist}(\zeta,\mathcal{Z}):=\inf\{\|\zeta-z\|,z\in\mathcal{Z}\}$.
\section{Problem statement}
\label{sec:problem_statement}
    \subsection{Stochastic system and probabilistic constraints}
	Consider the following discrete-time linear system
	\begin{equation}
	\left\{\begin{array}{l}
	x_{t+1}=Ax_t+Bu_t+Fw_t \quad t\geq 0\\
	y_{t}=Cx_t+v_t
	\end{array}\right .
	\label{eq:model}\end{equation}	
	where
	$x_t\in \mathbb{R}^n$
	is the state,
	$u_t\in\mathbb{R}^m$
	is the input,
	$y_t\in\mathbb{R}^p$
	is the measured output and
	$w_t\in\mathbb{R}^{n_w}, v_t\in\mathbb{R}^{p}$
	are two independent, zero-mean, white noises with covariance matrices $W\succeq 0$ and $V\succ0$, respectively, and a-priori unbounded support. The pair $(A,C)$ is assumed to be observable, and the pairs $(A,B)$ and $(A,\tilde{F})$ are reachable, where matrix $\tilde{F}$ satisfies $\tilde{F}\tilde{F}^T=FWF^T$.\\
	Polytopic constraints on the state and input variables of system \eqref{eq:model} are imposed in a probabilistic way, i.e., it is required that, for all $t\geq 0$
	\begin{align}\label{eq:prob_constraint_state}
	\mathbb{P}\{b_r^Tx_{t}\geq x^{ max}_r\}&\leq p_r^x \quad r=1,\dots, n_r\\
    \label{eq:prob_constraint_input}
	\mathbb{P}\{c_s^Tu_{t}\geq u^{ max}_s\}&\leq p_s^u \quad s=1,\dots, n_s
	\end{align}
	where $\mathbb{P}(\phi)$ denotes the probability of $\phi$, $b_r$, $c_s$ are constant vectors, $x_r^{ max}$, $u_s^{ max}$  are bounds for the state and control variables, and $p^x_r, p^u_s$ are design parameters.
	It is also assumed that the set of relations $b_r^Tx\leq x^{ max}_r$, $r=1,\dots, n_r$ (respectively, $c_s^Tu\leq u^{ max}_s$, $s=1,\dots, n_s$), defines a convex set $\mathbb{X}$ (respectively, $\mathbb{U}$) containing the origin in its interior.
\subsection{Regulator structure}
For system~\eqref{eq:model}, we want to design a standard regulation scheme made by the state observer
\begin{equation}\label{eq:observer1}
	 \hat{x}_{t+1}=A\hat{x}_{t}+Bu_t+L_{t}(y_t-C\hat{x}_{t})
\end{equation}	
coupled with the feedback control law
\begin{equation}\label{eq:fb_control_law_ideal}
	u_{t}=\bar{u}_{t}-K_{t}(\hat{x}_t-\bar{x}_{t})
\end{equation}
where $\bar{x}$ is the state of the nominal model
\begin{equation}\label{eq:mean_value_evolution_freeIC}
	\bar{x}_{t+1}=A\bar{x}_{t}+B\bar{u}_{t}
\end{equation}
In \eqref{eq:observer1}, \eqref{eq:fb_control_law_ideal}, the feedforward term $\bar{u}_{t}$ and the gains $L_{t}$, $K_{t}$ are design parameters to be selected to guarantee convergence properties and the fulfillment of the probabilistic constraints \eqref{eq:prob_constraint_state}, \eqref{eq:prob_constraint_input}.\\
Letting
\begin{subequations}
\label{eq:errors}
	\begin{align}
    e_{t}&=x_{t}-\hat{x}_{t}\label{eq:obs_error}\\
    \varepsilon_{t}&=\hat{x}_{t}-\bar{x}_{t}\label{est_error}
	\end{align}	
\end{subequations}
	from \eqref{eq:errors} we obtain that
	\begin{equation}\label{eq:error2}
	\delta x_{t}=x_{t}-\bar{x}_{t}=
	e_{t}+\varepsilon_{t}
	\end{equation}
	Define also the vector
    $\sigma_{t}=\begin{bmatrix}e_{t}^T& \varepsilon_{t}^T\end{bmatrix}^T$
    whose dynamics, according to \eqref{eq:model}-\eqref{eq:errors}, is described by
	\begin{equation}\label{eq:error_matrix}
\begin{array}{ll}
	\sigma_{t+1}=&\Phi_{t} \sigma_{t}+\Psi_{t}
	 \begin{bmatrix}w_{t}\\v_{t}\end{bmatrix}\end{array}
	\end{equation}
where $$\Phi_{t}=
 \begin{bmatrix}A-L_{t}C&0\\L_{t}C&A-BK_{t}\end{bmatrix},\,\Psi_{t}=\begin{bmatrix}F&-L_{t}\\0&L_{t}\end{bmatrix}$$
 In the following it is assumed that, by a proper initialization, i.e. $\mathbb{E}\left\{\sigma_{0}\right\}=0$, and recalling that the noises $v$ and $w$ are zero mean, the enlarged state $\sigma_{t}$ of system \eqref{eq:error_matrix} is zero-mean, so that $\bar{x}_{t}=\mathbb{E}\{x_{t}\}$. Then, denoting by
$\Sigma_{t}=\mathbb{E}\left\{
	\sigma_{t}\sigma_{t}^T
	\right\}$ and by $\Omega=\mathrm{diag}(W,V)$ the covariance matrices of $\sigma_{t}$ and $[w_{t}^T\, v_{t}^T]^T$ respectively, the evolution of $\Sigma_{t}$ is governed by
	\begin{align}\label{eq:variance_evolution_error}
	 &\Sigma_{t+1}=\Phi_{t}\Sigma_{t}\Phi_{t}^T+\Psi_{t}\Omega\Psi_{t}^T
	\end{align}
	By definition, also the variable $\delta x_{t}$ defined by \eqref{eq:error2} is zero mean and its covariance matrix $X_{t}$ can be derived from $\Sigma_{t}$ as follows
	\begin{align}\label{eq:variance_evolution_state}
	X_{t}=\mathbb{E}\left\{\delta x_{t} \delta x_{t}^T\right\}
	=\begin{bmatrix}I&I\end{bmatrix}\Sigma_{t}
	\begin{bmatrix}I\\I\end{bmatrix}
	\end{align}
    Finally, letting $\delta u_{t}=u_{t}-\bar{u}_{t}=-K_{t}(\hat{x}_{t}-\bar{x}_{t})$, one has $\mathbb{E}\left\{\delta u_{t}\right\}=0$ and also the covariance matrix
    $U_{t}=\mathbb{E}\left\{\delta u_{t} \delta u_{t}^T\right\}$ can be obtained from $\Sigma_{t}$ as follows
	\begin{align}\label{eq:variance_evolution_input}
	U_{t}=&\mathbb{E}\left\{K_{t}\varepsilon_{t} \varepsilon_{t}^TK_{t}^T\right\}=\begin{bmatrix}0&K_{t}\end{bmatrix}\Sigma_{t}
	\begin{bmatrix}0\\K_{t}^T\end{bmatrix}
	\end{align}
\subsection{Reformulation of the probabilistic constraints}
To set up a suitable control algorithm for the design of $\bar{u}_{t}$, $L_{t}$, $K_{t}$, the probabilistic constraints  \eqref{eq:prob_constraint_state} and \eqref{eq:prob_constraint_input} are now reformulated as deterministic ones at the price of suitable tightening. To this end, consider, in general, a random variable $z$ with mean value $\bar{z}=\mathbb{E}\{z\}$, variance $Z=\mathbb{E}\{(z-\bar{z})(z-\bar{z})^T\}$, and the chance-constraint
\begin{equation}\mathbb{P}\{h^T z\geq z^{ max}\}\leq p\label{eq:prob_constraint_general}\end{equation}
The following result, based on the Chebyshev - Cantelli inequality \cite{Cantelli}, has been proven in~\cite{Pala}.
%
%can be reformulated as a suitable constraint, featuring its , i.e.,
%
%where $f(p)$ is a function of the probability $p$. Importantly, in case the noise distribution is unknown, constraints will be tightened solely based on the second-order description of the stochastic process $z$ and resorting of the Cantelli inequality. More specifically, the following proposition holds.
%
%
\begin{proposition}
\label{prop:Cantelli}
Letting $f(p)=\sqrt{(1-p)/{p}}$, constraint \eqref{eq:prob_constraint_general}
is verified if
\begin{equation}
h^T\bar{z}\leq z^{ max}-\sqrt{h^T Z h}\,f(p)
\label{eq:Cantelli_propGen}
\end{equation}
\end{proposition}
Note that this result can be proved without introducing any specific assumption on the distribution of~$z$. If, on the other hand, $z$ can be assumed to be normally distributed, less conservative constraints can be obtained, as stated in the following result.
\begin{proposition}
\label{prop:gen_distr}
Assume that $z$ is normally distributed. Then, constraint \eqref{eq:prob_constraint_general}
is verified if \eqref{eq:Cantelli_propGen} holds with $f(p)=\mathcal{N}^{-1}(1-p)$ where $\mathcal{N}$ is the cumulative probability function of a Gaussian variable with zero mean and unitary variance.
\end{proposition}
In Propositions \ref{prop:Cantelli} and \ref{prop:gen_distr}, the function $f(p)$ represents the level of constraint tightening on the mean value of $z$ needed to meet the probabilistic constraint \eqref{eq:prob_constraint_general}. In case of unknown distribution (Proposition \ref{prop:Cantelli}) the values of $f(p)$ are significantly smaller with respect to the Gaussian case (e.g., about an order of magnitude in the range $(0.1, 0.4)$). Similar results can be derived in case of different distributions (e.g., homogeneous).\\
%    \begin{figure}
%    \centering
%    \includegraphics[width=8cm]{Gauss_gen}
%    \caption{Comparison of the values of $f(p)$ in case of unknown distribution (solid line) and in case of Gaussian distribution (dashed line).}
%    \label{fig:Gauss_gen}
%    \end{figure}
%
%
%
In view of Propositions \ref{prop:Cantelli} and \ref{prop:gen_distr}, the chance-constraints \eqref{eq:prob_constraint_state}-\eqref{eq:prob_constraint_input} are verified provided that the following (deterministic) inequalities are satisfied.
\begin{subequations}
\begin{align}
b_r^T\bar{x}_{t}&\leq x_r^{max}-\sqrt{b_r^T X_{t} b_r}f(p^x_r)\label{eq:Cantelli_ineqs_state}\\
c_s^T\bar{u}_{t}&\leq u_s^{max}-\sqrt{c_s^T U_{t} c_s}f(p_s^u)\label{eq:Cantelli_ineqs_input}
\end{align}
\label{eq:Cantelli_ineqs}
\end{subequations}
If the support of the noise terms $w_k$ and $v_k$ is unbounded, the definition of state and control constraints in probabilistic terms is the only way to state feasible control problems. In case of bounded noises the comparison, in terms of conservativeness between the probabilistic framework and the deterministic one, is discussed in the example in Section \ref{app:example_constrs}.
	\section{MPC algorithm: formulation and properties}
	\label{MPC}
	To formally state the MPC algorithm for the computation of the regulator parameters $\bar{u}_{t}$, $L_{t}$, $K_{t}$, the following notation will be adopted: given a variable $z$ or a matrix $Z$, at any time step $t$ we will denote by $z_{t+k}$ and $Z_{t+k}$, $k\geq 0$, their generic values in the future, while $z_{t+k|t}$ and $Z_{t+k|t}$ will represent their specific values computed based on the knowledge (e.g., measurements) available at time $t$.\\
The main ingredients of the optimization problem are now introduced.
\subsection{Cost function}
Assume to be at time $t$ and denote by $\bar{u}_{t,\dots, t+N-1}=\{\bar{u}_t, \dots, \bar{u}_{t+N-1}\}$ the nominal input sequence over a future prediction horizon of length $N$. Moreover, define by $K_{t,\dots, t+N-1}=\{K_t, \dots, K_{t+N-1}\}$, $L_{t,\dots, t+N-1}=\{L_t, \dots, L_{t+N-1}\}$ the sequences of the future control and observer gains, and recall that the covariance $\Sigma_{t+k}=\mathbb{E}\left\{\sigma_{t+k}\sigma_{t+k}^T\right\}$ evolves, starting from $\Sigma_{t}$, according to \eqref{eq:variance_evolution_error}.\\
The cost function to be minimized is the sum of two components, the first one ($J_{m}$) accounts for the expected values of the future nominal inputs and states, while the second one ($J_{v}$) is related to the variances of the future errors $e$, $\varepsilon$, and of the future inputs. Specifically, the overall performance index is
	 \begin{align}\label{eq:JTOT}%\label{eq:divided_cost_function}
	J=J_m(\bar{x}_t,\bar{u}_{t,\dots, t+N-1})+J_v(\Sigma_t,K_{t,\dots, t+N-1},L_{t,\dots, t+N-1})
	\end{align}
	where
	\begin{align}
	&J_m=\sum_{i=t}^{t+N-1} \| \bar{x}_{i}\|_{Q}^2+\| \bar{u}_{i}\|_{R}^2+\| \bar{x}_{t+N}\|_{S}^2\label{eq:mean_cost_function}\\
	&J_v=\mathbb{E}\left\{\sum\limits_{i=t}^{t+N-1} \| x_i-\hat{x}_i \|_{Q_L}^2+\| x_{t+N}-\hat{x}_{t+N} \|_{S_{L}}^2 \right\}+\nonumber\\ &\mathbb{E}\left\{\sum\limits_{i=t}^{t+N-1} \| \hat{x}_i-\bar{x}_i \|_Q^2+\| u_i-\bar{u}_i \|_R^2+\| \hat{x}_{t+N}-\bar{x}_{t+N} \|_{S}^2 \right\} \label{eq:variance_cost_function1}\end{align}
where the positive definite and symmetric weights $Q$, $Q_L$, $S$, and $S_L$ must satisfy the following inequality
	\begin{equation}\label{eq:Lyap_S} Q_{T}-S_{T}+\Phi^TS_{T}\Phi\preceq 	 0 \end{equation}
	where $$\Phi=\begin{bmatrix}A-\bar{L}C&0\\\bar{L}C&A-B\bar{K}\end{bmatrix}$$
$Q_{T}=\mathrm{diag}(Q_{L},Q+\bar{K}^TR\bar{K})$, $S_{T}=\mathrm{diag}(S_{L},S)$,
and $\bar{K}$, $\bar{L}$ must be chosen to guarantee that $\Phi$ is asymptotically stable.\\
By means of standard computations, it is possible to write the cost \eqref{eq:variance_cost_function1} as follows
\begin{equation}\label{eq:variance_cost_function}
J_v=\sum_{i=t}^{t+N-1} \mathrm{tr}(Q_{T} \Sigma_{i})+ \mathrm{tr} (S_{T} \Sigma_{t+N})
\end{equation}
From \eqref{eq:JTOT}-\eqref{eq:variance_cost_function1}, it is apparent that the goal is twofold: to drive the mean $\bar{x}$ to zero by acting on the nominal input component $\bar{u}_{t,\dots, t+N-1}$ and to minimize the variance of $\Sigma$ by acting on the gains $K_{t,\dots, t+N-1}$ and $L_{t,\dots, t+N-1}$. In addition, also the pair $(\bar{x}_{t},\Sigma_t)$ must be considered as an additional argument of the MPC optimization, as later discussed, to guarantee recursive feasibility.
\subsection{Terminal constraints}
	As usual in stabilizing MPC, see e.g. \cite{Mayne00}, some terminal constraints must be included into the problem formulation. In our setup, the mean $\bar{x}_{t+N}$ and the variance $\Sigma_{t+N}$ at the end of the prediction horizon must satisfy
	\begin{align}
	 \bar{x}_{t+N}&\in\bar{\mathbb{X}}_F\label{eq:term_constraint_mean}\\
	\Sigma_{t+N}&\preceq \bar{\Sigma}\label{eq:term_constraint_variance}
	\end{align}
where $\bar{\mathbb{X}}_F$ is a positively invariant set (see \cite{Gilbert}) such that
	\begin{align}\label{eq:inv_terminal}
	(A-B\bar{K})\bar{x}&\in\bar{\mathbb{X}}_F  \quad &\forall \bar{x}\in \bar{\mathbb{X}}_F
	\end{align}
while $\bar{\Sigma}$ is the steady-state solution of the Lyapunov equation \eqref{eq:variance_evolution_error}, i.e.,
\begin{align}
\bar{\Sigma}=&
    \Phi
	 \bar{\Sigma}
	 \Phi^T+\Psi\bar{\Omega}\Psi^T
\label{eq:Riccati_1}
\end{align}
where
$\Psi=\begin{bmatrix}F&-\bar{L}\\0  &\bar{L} \end{bmatrix}$
and $\bar{\Omega}=\mathrm{diag}(\bar{W},\bar{V})$ is built by considering (arbitrary) noise variances $\bar{W}\succeq W$ and $\bar{V}\succeq V$. %, and assuming constant gains $\bar{K}$ and $\bar{L}$.\\
In addition, and consistently with \eqref{eq:Cantelli_ineqs}, the following coupling conditions, must be verified.
	\begin{subequations}
    \label{eq:linear_constraint_finalc}
	\begin{align}
	b_r^T\bar{x}&\leq x_r^{max}-\sqrt{b_r^T \bar{X} b_r}f(p^x_r) \label{eq:linear_constraint_state_finalc}\\
	-c_s^T\bar{K}\bar{x}&\leq u_s^{max}-\sqrt{c_s^T \bar{U} c_s}f(p_s^u) \label{eq:linear_constraint_input_finalc}
	\end{align}
	\end{subequations}
%
%
%
%	\begin{subequations}
%	\begin{align}
%	b_r^T\bar{x}&\leq (1-0.5\alpha^x)x_r^{ max}-\frac{1}{2\alpha^x x_r^{ max}}b_r^T\bar{X}b_r f(p_r^x)^2\label{eq:linear_constraint_state_final}\\
%	-c_s^T\bar{K}\bar{x}&\leq (1-0.5\alpha^u)u_s^{ max}-\frac{1}{2\alpha^u u_s^{ max}}c_s^T\bar{U}c_s f(p_s^u)^2 \label{eq:linear_constraint_input_final}
%	\end{align}
%	\end{subequations}
	for all $r=1,\dots, n_r$, $s=1,\dots, n_s$, and for all $\bar{x}\in\bar{\mathbb{X}}_F$, where
	\begin{subequations}
    \label{eq:bar_def}
    \begin{align}
    \bar{X}&=\begin{bmatrix}I&I\end{bmatrix}\bar{\Sigma}\begin{bmatrix}I\\I\end{bmatrix}\label{eq:Xbar_def}\\
    \bar{U}&=\begin{bmatrix}0&\bar{K}\end{bmatrix}\bar{\Sigma}\label{eq:Ubar_def}
	\begin{bmatrix}0\\\bar{K}^T\end{bmatrix}
	\end{align}\end{subequations}
It is worth remarking that the choice of $\bar{\Omega}$ is subject to a tradeoff. In fact, large variances $\bar{W}$ and $\bar{V}$ result in large $\bar{\Sigma}$ (and, in view of \eqref{eq:bar_def}, large $\bar{X}$ and $\bar{U}$). This enlarges the terminal constraint \eqref{eq:term_constraint_variance} but, on the other hand, reduces the size of the terminal set $\mathbb{X}_F$ compatible with \eqref{eq:linear_constraint_finalc}.
	\subsection{Statement of the stochastic MPC (S-MPC) problem}
The formulation of the main S-MPC problem requires a  preliminary discussion concerning the initialization. In principle, and in order to use the most recent information available on the state, at any time instant it would be natural to set the current value of the nominal state $\bar{x}_{t|t}$ to $\hat x_{t}$ and the covariance $\Sigma_{t|t}$ to $\mathrm{diag}(\Sigma_{11,t|t-1},0)$, where $\Sigma_{11,t|t-1}$ is the covariance of state prediction error $e$ obtained using the observer~\eqref{eq:observer1}. However, since we do not exclude the possibility of unbounded disturbances, in some cases this choice could lead to infeasible optimization problems. On the other hand, and in view of the terminal constraints \eqref{eq:term_constraint_mean}, \eqref{eq:term_constraint_variance}, it is quite easy to see that recursive feasibility is guaranteed provided that $\bar{x}$ is updated according to the prediction equation \eqref{eq:mean_value_evolution_freeIC}, which corresponds to the variance update given by \eqref{eq:variance_evolution_error}. These considerations motivate the choice of accounting for the initial conditions $(\bar{x}_t,\Sigma_t)$ as free variables, which will selected by the control algorithm according to the following alternative strategies.\\
	\textbf{Strategy 1} Reset of the initial state: $\bar{x}_{t|t}=\hat x_{t}$, $\Sigma_{t|t}=\mathrm{diag}(\Sigma_{11,t|t-1},0)$.\\
	\textbf{Strategy 2} Prediction: $\bar{x}_{t|t}=\bar{x}_{t|t-1}$, $\Sigma_{t|t}=\Sigma_{t|t-1}$.\\
The S-MPC problem can now be stated.\\\\
	\textbf{S-MPC problem: }at any time instant $t$ solve

    $$\min_{\bar{x}_{t},\Sigma_{t},\bar{u}_{t, \dots, t+N-1},
	K_{t,\dots, t+N-1},L_{t,\dots, t+N-1}} J$$
where $J$ is defined in \eqref{eq:JTOT}, \eqref{eq:mean_cost_function}, \eqref{eq:variance_cost_function1}, subject to
\begin{itemize}
  \item[-] the dynamics~\eqref{eq:mean_value_evolution_freeIC} and~\eqref{eq:variance_evolution_error};
  \item[-] constraints~\eqref{eq:Cantelli_ineqs} for all $k=0,\dots,N-1$;
  \item[-] the initialization constraint, corresponding to the choice between Strategies 1 and 2, i.e.,
	\begin{equation}\label{eq:reset_constraint}
	  (\bar{x}_{t},\Sigma_{t})\in\{(\hat x_t,\mathrm{diag}(\Sigma_{11,t|t-1},0)),(\bar{x}_{t|t-1},\Sigma_{t|t-1})\}
	\end{equation}
    \item[-] the terminal constraints~\eqref{eq:term_constraint_mean}, \eqref{eq:term_constraint_variance}.\hfill$\square$
\end{itemize}
	Denoting by
	$\bar{u}_{t,\dots, t+N-1|t}=\{\bar{u}_{t|t},\dots, \bar{u}_{t+N-1|t}\}$,
	$K_{t,\dots, t+N-1|t}=\{K_{t|t},\dots,$\break
$K_{t+N-1|t}\}$,
	$L_{t,\dots, t+N-1|t}=$
$\{L_{t|t},\dots, L_{t+N-1|t}\}$, and
($\bar{x}_{t|t},\Sigma_{t|t}$)
	the optimal solution of the S-MPC problem, the feedback control law actually used is then given by~\eqref{eq:fb_control_law_ideal} with
$\bar{u}_{t}=\bar{u}_{t|t}$, $K_{t}=K_{t|t}$, and the state observation evolves as in~\eqref{eq:observer1} with $L_{t}=L_{t|t}$.\\
	We define the S-MPC problem feasibility set as
\begin{center}
	$\Xi:=\{(\bar{x}_0,\Sigma_0):\exists \bar{u}_{0,\dots, N-1},K_{0,\dots, N-1},L_{0,\dots, N-1}$
such that~\eqref{eq:mean_value_evolution_freeIC},~\eqref{eq:variance_evolution_error}, and \eqref{eq:Cantelli_ineqs} hold for all $k=0,\dots,N-1$ and \eqref{eq:term_constraint_mean}, \eqref{eq:term_constraint_variance} are verified\}\end{center}
%Note that, in view of the compactness of $\mathbb{X}$, see \eqref{eq:prob_constraint_state}, the set $\Xi$ results to be compact.\\
Some comments are in order.
	\setlength{\leftmargini}{0.5em}
	\begin{itemize}
	\item[-]At the initial time $t=0$, the algorithm must be initialized by setting $\bar{x}_{0|0}=\hat{x}_{0}$ and $\Sigma_{0|0}=\mathrm{diag}(\Sigma_{11,0},0)$. In view of this, feasibility at time $t=0$ amounts to $(\hat{x}_0,\Sigma_{0|0})\in\Xi$.
%	\item[-] According to the problem statement, feasibility of S-MPC at time $t>0$ amounts to $\{(\hat{x}_t,$diag$(\Sigma_{11,t},0)),(\bar{x}_{t|t-1},\Sigma_{t|t-1})\}\bigcap\Xi\neq\emptyset$.
	\item[-] The binary choice between Strategies 1 and 2 requires to solve at any time instant two optimization problems. However, the following sequential procedure can be adopted to reduce the average overall computational burden: the optimization problem corresponding to Strategy 1 is first solved and, if it is infeasible, Strategy 2 must be used, otherwise Strategy 2 must be solved and adopted. On the contrary, if it is feasible, it is possible to compare the resulting value of the optimal cost function with the value of the cost using the sequences $\{\bar{u}_{t|t-1},\dots, \bar{u}_{t+N-2|t-1}, -\bar{K}\bar{x}_{t+N-1|t}\}$, $\{K_{t|t-1},\dots, K_{t+N-2|t-1}, \bar{K}\}$, $\{L_{t|t-1},\dots,$ $L_{t+N-2|t-1}, \bar{L}\}$. If the optimal cost with Strategy 1 is lower, Strategy 1 can be used without solving the MPC problem for Strategy 2. This does not guarantee optimality, but the convergence properties of the method stated in the result below are recovered and the computational effort is reduced.
	\end{itemize}
	Now we are in the position to state the main result concerning the convergence properties of the algorithm.
	\begin{thm}\label{thm:main}
    If, at $t=0$, the S-MPC problem admits a solution, the optimization problem is recursively feasible and the state and input probabilistic constraints \eqref{eq:prob_constraint_state} and \eqref{eq:prob_constraint_input} are satisfied for all $t\geq 0$. Furthermore, if there exists $\rho\in(0,1)$ such that the noise variance $\Omega$ verifies
	\begin{align}
	 \frac{(N+\frac{\beta}{\alpha})}{\alpha}\mathrm{tr}(S_T\Psi \Omega \Psi^T)&<\min(\rho\bar{\sigma}^2,\rho\lambda_{min}(\bar{\Sigma}))\label{eq:cons_conds_on_W}
	\end{align}
	where $\bar{\sigma}$ is the maximum radius of a ball, centered at the origin, included in $\bar{\mathbb{X}}_F$, and
    \begin{subequations}
        \begin{align}
        \alpha&=\min\{\lambda_{min}(Q),\mathrm{tr}\{Q^{-1}+Q_L^{-1}\}^{-1}\}\\
        \beta&=\max\{\lambda_{max}(S),\mathrm{tr}\{S_T\}\}
        \end{align}
        \label{eq:lambda_def}
    \end{subequations}
	then, as $t\rightarrow +\infty$\\
    \begin{align}\mathrm{dist}(\|\bar{x}_t\|^2+\mathrm{tr}\{\Sigma_{t|t}\},[0,\frac{1}{\alpha}(N+\frac{\beta}{\alpha})\,\mathrm{tr}(S_T\Psi \Omega \Psi^T)])\rightarrow 0\label{eq:thm_stat}\end{align}\hfill$\square$
    \end{thm}
    Note that, as expected, for smaller and smaller values of $\Omega$, also the asymptotic values of $\|\bar{x}_t\|$ and $\mathrm{tr}\{\Sigma_{t|t}\}$  tend to zero.
\section{Implementation issues}
\label{sec:num_implementation}
The main difficulty in the solution to the S-MPC problem is due to the non linear constraints \eqref{eq:Cantelli_ineqs} and to the non linear dependence of the covariance evolution,  see~\eqref{eq:variance_evolution_error}, on $K_{t,\dots, t+N-1},L_{t,\dots, t+N-1}$. This second problem can be prevented in the state feedback case, see \cite{FGMS13_CDC}, where a reformulation based on linear matrix inequalities (LMIs) can be readily obtained. In the output feedback case here considered, two possible solutions are described in the following.\\
Also, in Section \ref{sec:inputs} we briefly describe some possible solutions for coping with the presence of additive deterministic constraints on the input variables $u_t$.
\subsection{Approximation of S-MPC for allowing a solution with LMIs}
A solution, based on an approximation of S-MPC characterized by linear constraints solely, is now presented.
First define $A^D=\sqrt{2}A,B^D=\sqrt{2}B,C^D=\sqrt{2}C$, and $V^D=2V$ and let the auxiliary gain matrices $\bar{K}$ and $\bar{L}$ be selected according to the following assumption.
\begin{assumption}
\label{ass:KandL}
The gains $\bar{K}$ and $\bar{L}$ are computed as the steady-state gains of the LQG regulator for the system $(A^D,B^D,C^D)$, with state and control weights $Q$ and $R$, and noise covariances $\bar{W}\succeq W$ and $\bar{V}\succeq V^D$.
\end{assumption}
Note that, if a gain matrix $\bar{K}$ (respectively $\bar{L}$) is stabilizing for $(A^D-B^D\bar{K})=\sqrt{2}(A-B\bar{K})$ (respectively $(A^D-\bar{L}C^D)=\sqrt{2}(A-\bar{L}C)$), it is also stabilizing for $(A-B\bar{K})$ (respectively $(A-\bar{L}C)$), i.e., for the original system. The following preliminary result can be stated.
\begin{lemma}
\label{lemma:bound_var_main}
Define $A^D_{L_t}=A^D-L_tC^D$, $A^D_{K_t}=A^D-B^DK_t$, the block diagonal matrix $\Sigma^D_t=\mathrm{diag}(\Sigma_{11,t}^D,\Sigma_{22,t}^D)$, $\Sigma_{11,t}^D\in \mathbb{R}^{n\times n}$, $\Sigma_{22,t}^D\in \mathbb{R}^{n\times n}$  and the update equations	
\begin{subequations}
\begin{align}
	 \Sigma_{11,t+1}^D=&A^D_{L_t}\Sigma_{11,t}^D(A^D_{L_t})^T+FWF^T+L_t V^D L_t^T\label{eq:sigmaD1}\\	 \Sigma_{22,t+1}^D=&A^D_{K_t}\Sigma_{22,t}^D(A^D_{K_t})^T+L_tC^D\Sigma_{11,t}^DC^{D\ T}L_t^T\nonumber\\
&+L_t V^D L_t^T\label{eq:sigmaD2}
\end{align}
\label{eq:SigmaDupdate}
\end{subequations}
Then\\
I) $\Sigma^D_t\succeq \Sigma_t$ implies that $\Sigma^D_{t+1}=\mathrm{diag}(\Sigma_{11,t+1}^D,\Sigma_{22,t+1}^D)\succeq \Sigma_{t+1}$.\\
II) We can rewrite as LMIs the following inequalities
\begin{subequations}
\begin{align} \Sigma_{11,t+1}^D\succeq &A^D_{L_t}\Sigma_{11,t}^D(A^D_{L_t})^T+FWF^T+L_t V^D L_t^T\label{eq:sigmaD1LMI}\\ \Sigma_{22,t+1}^D \succeq & A^D_{K_t}\Sigma_{22,t}^D(A^D_{K_t})^T+L_tC^D\Sigma_{11,t}^DC^{D\ T}L_t^T\nonumber\\
&+L_t V^D L_t^T\label{eq:sigmaD2LMI}
\end{align}
\label{eq:SigmaDupdateLMI}
\end{subequations}
\end{lemma}
%
%\begin{lemma}
%\label{lemma:bound_var_main}
%Define matrix $\Sigma^D_t$ in such a way that $\Sigma^D_t=$diag$(\Sigma_{11,t}^D,\Sigma_{22,t}^D)$ and that $\Sigma^D_t\succeq \Sigma_t$. If we define $\Sigma^D_{t+1}=$diag$(\Sigma_{11,t+1}^D,\Sigma_{22,t+1}^D)$, where	
%\begin{subequations}
%\begin{align}
%	 \Sigma_{11,t+1}^D&=(A^D-L_tC^D)\Sigma_{11,t}^D(A^D-L_tC^D)^T+\nonumber\\
%		&FWF^T+L_t V^D L_t^T\label{eq:sigmaD1}\\
%	 \Sigma_{22,t+1}^D&=(A^D-B^DK_t)\Sigma_{22,t}^D(A^D-B^DK_t)^T+\nonumber\\
%		&L_tC^D\Sigma_{11,t}^DC^{D\ T}L_t^T+L_t V^D L_t^T\label{eq:sigmaD2}
%\end{align}
%\label{eq:SigmaDupdate}
%\end{subequations}
%%
%and where $A^D=\sqrt{2}A,B^D=\sqrt{2}B,C^D=\sqrt{2}C$ and $V^D=2V$,
%then $$\Sigma^D_{t+1}\succeq \Sigma_{t+1}$$
%\end{lemma}
%
Based on Lemma~\ref{lemma:bound_var_main}-II, we can reformulate the original problem so that the covariance matrix $\Sigma^{D}$ is used instead of $\Sigma$. Accordingly, the update equation~\eqref{eq:variance_evolution_error} is replaced by \eqref{eq:SigmaDupdate} and S-MPC problem is recast as an LMI one (see Appendix \ref{app:LMI}).\\
The inequalities \eqref{eq:Cantelli_ineqs} have a nonlinear dependence on the covariance matrices $X_{t}$ and $U_{t}$. It is possible to prove that \eqref{eq:Cantelli_ineqs} are satisfied if
\begin{subequations}\label{eq:Cantelli_ineqs_lin}
	\begin{align}
	b_r^T\bar{x}_{t}&\leq
	(1-0.5\alpha_x)x_r^{ max}-\frac{b_r^T X_{t}b_r}{2\alpha_x x_r^{ max}} f(p_r^x)^2 \label{eq:linear_constraint_state}\\
	c_s^T\bar{u}_{t}&\leq (1-0.5\alpha_u)u_s^{ max}-\frac{c_s^TU_{t}c_s}{2\alpha_u u_s^{ max}} f(p_s^u)^2\label{eq:linear_constraint_input}
\end{align}\end{subequations}
	with $r=1,\dots, n_r$ and $s=1,\dots, n_s$, where $\alpha_x\in(0, 1]$ and $\alpha_u\in(0, 1]$ are free design parameters. Also, note that
$X_{t}\preceq\begin{bmatrix}
	I&I
	\end{bmatrix}\Sigma_{t}^D\begin{bmatrix}
	I\\I
	\end{bmatrix}= \Sigma_{11,t}^D+\Sigma_{22,t}^D$ and that $U_{t}\preceq\begin{bmatrix}
	0&K_{t}
	\end{bmatrix}\Sigma_{t}^D\begin{bmatrix}
	0\\K_{t}^T
	\end{bmatrix}=K_{t}\Sigma_{22,t}^DK_{t}^T$
so that, defining $X_{t}^D=\Sigma_{11,t}^D+ \Sigma_{22,t}^D$ and $U_{t}^D=K_{t}\Sigma_{22,t}^DK_{t}^T$,~\eqref{eq:Cantelli_ineqs_lin} can be written as follows
\begin{subequations}\label{eq:Cantelli_ineqs_linL}
	\begin{align}
	b_r^T\bar{x}_{t}&\leq
	(1-0.5\alpha_x)x_r^{ max}-\frac{b_r^TX_{t}^D b_r}{2\alpha_x x_r^{ max}} f(p_r^x)^2\label{eq:linear_constraint_stateL}\\
	c_s^T\bar{u}_{t}&\leq (1-0.5\alpha_u)u_s^{ max}-\frac{c_s^TU_{t}^D c_s}{2\alpha_u u_s^{ max}} f(p_s^u)^2\label{eq:linear_constraint_inputL}
\end{align}\end{subequations}
Note that the reformulation of~\eqref{eq:Cantelli_ineqs} into \eqref{eq:Cantelli_ineqs_linL} has been performed at the price of additional constraint tightening. For example, on the right hand side of \eqref{eq:linear_constraint_stateL}, $x_r^{max}$ is replaced by $(1-0.5\alpha^x)x_r^{max}$, which significantly reduces the size of the constraint set. Note that parameter $\alpha^x$ cannot be reduced at will, since it also appears at the denominator in the second additive term.\\
In view of Assumption~\ref{ass:KandL} and resorting to the separation principle, it is possible to show \cite{glad2000control} that the solution $\bar{\Sigma}^D$ to the steady-state equation
\begin{align}
\bar{\Sigma}^D=&
    \Phi^D
	 \bar{\Sigma}^D
	 (\Phi^D)^T+\Psi\bar{\Omega}\Psi^T
\label{eq:Riccati_1L}
\end{align}
is block diagonal, i.e., $\bar{\Sigma}^D=\mathrm{diag}(\bar{\Sigma}^D_{11},\bar{\Sigma}_{22}^D)$,
where
$$\Phi^D=\begin{bmatrix}A^D-\bar{L}C^D&0\\\bar{L}C^D&A^D-B^D\bar{K}\end{bmatrix}$$
The terminal constraint \eqref{eq:term_constraint_variance}, must be transformed into $\Sigma_{t+N}^D\preceq \bar{\Sigma}^D$,
which corresponds to setting
\begin{equation}
\begin{array}{lcllcl}
	\Sigma_{11,t+N}^D&\preceq \bar{\Sigma}^D_{11},\quad&
    \Sigma_{22,t+N}^D&\preceq \bar{\Sigma}^D_{22}
\end{array}
\label{eq:term_constraints_varianceL}
\end{equation}
Defining $\bar{X}^D=\bar{\Sigma}^D_{11}+\bar{\Sigma}^D_{22}$ and $\bar{U}^D=\bar{K}\bar{\Sigma}^D_{22}\bar{K}^T$, the terminal set condition \eqref{eq:linear_constraint_finalc} must now be reformulated as
\begin{subequations}
	\begin{align}
	b_r^T\bar{x}&\leq (1-0.5\alpha^x)x_r^{ max}-\frac{b_r^T\bar{X}^D b_r}{2\alpha^x x_r^{ max}} f(p_r^x)^2 \label{eq:linear_constraint_state_finalL}\\
	-c_s^T\bar{K}\bar{x}&\leq (1-0.5\alpha^u)u_s^{ max}-\frac{c_s^T \bar{U}^Dc_s}{2\alpha^u u_s^{ max}} f(p_s^u)^2 \label{eq:linear_constraint_input_finalL}
	\end{align}
\end{subequations}
for all $r=1,\dots, n_r$, $s=1,\dots, n_s$, and for all $\bar{x}\in\bar{\mathbb{X}}_F$.\\
Also $J_v$ must be reformulated. Indeed
\begin{equation}
	\begin{array}{ll}
	J_v\leq J_v^D&=\sum\limits_{i=t}^{t+N-1} \mathrm{tr}\left\{Q_L\Sigma_{11,i}^D+Q\Sigma_{22,i}^D+RK_i\Sigma_{22,i}^DK_i^T\right\}\\
&+\mathrm{tr}\left\{S_L\Sigma_{11,t+N}^D+S\Sigma_{22,t+N}^D\right\}
	\end{array}
\label{eq:variance_cost_functionL}
\end{equation}
where the terminal weights $S$ and $S_L$ must now satisfy the following Lyapunov-like inequalities
\begin{equation}
\begin{array}{l}
(\bar{A}^D_K)^T S \bar{A}^D_K-S+Q+\bar{K}^TR\bar{K}\preceq 0\\
(\bar{A}^D_L)^T S_L \bar{A}^D_L-S_L+Q_L+(C^D)^T\bar{L}^T S \bar{L}C^D\preceq 0
\end{array}
\label{eq:Lyap_S_L}
\end{equation}
where $\bar{A}^D_K=A^D-B^D\bar{K}$ and $\bar{A}^D_L=A^D-\bar{L}C^D$. It is now possible to formally state the S-MPCl problem.\\\\
\textbf{S-MPCl problem: }at any time instant $t$ solve
    $$\min_{\bar{x}_{t},\Sigma^D_{11,t},\Sigma^D_{22,t},\bar{u}_{t, \dots, t+N-1},K_{t,\dots, t+N-1},L_{t,\dots, t+N-1}} J$$
where $J$ is defined in \eqref{eq:JTOT}, \eqref{eq:mean_cost_function}, \eqref{eq:variance_cost_functionL}, subject to
\begin{itemize}
  \item[-] the dynamics~\eqref{eq:mean_value_evolution_freeIC} and~\eqref{eq:SigmaDupdate};
  \item[-] the linear constraints~\eqref{eq:Cantelli_ineqs_linL} for all $k=0,\dots,N-1$;
  \item[-] the initialization constraint, corresponding to the choice between Strategies 1 and 2, i.e., $(\bar{x}_{t},\Sigma^D_{11,t},\Sigma_{22,t}^D)\in\{(\hat x_t,\Sigma^D_{11,t|t-1},0),(\bar{x}_{t|t-1},\Sigma^D_{11,t|t-1},\Sigma_{22,t|t-1}^D)\}$
    \item[-] the terminal constraints~\eqref{eq:term_constraint_mean},~\eqref{eq:term_constraints_varianceL}.
\end{itemize}
\hfill$\square$\\
The following corollary follows from Theorem~\ref{thm:main}.
\begin{corollary}\label{cor:LMI_soluz}
If, at time $t=0$, the S-MPCl problem admits a solution, the optimization problem is recursively feasible and the state and input probabilistic constraints \eqref{eq:prob_constraint_state} and \eqref{eq:prob_constraint_input} are satisfied for all $t\geq 0$. Furthermore, if there exists $\rho\in(0,1)$ such that the noise variance $\Omega^D=\mathrm{diag}(W,V^D)$ verifies
\begin{align}
	 \frac{(N+\frac{\beta}{\alpha})}{\alpha}\mathrm{tr}(S_T\Psi \Omega^D \Psi^T)&<\min(\rho\bar{\sigma}^2,\rho\lambda_{min}(\bar{\Sigma}^D))\label{eq:cons_conds_on_WL}
\end{align}
then, as $t\rightarrow +\infty$
$$\mathrm{dist}(\|\bar{x}_t\|^2+\mathrm{tr}\{\Sigma_{t|t}^D\},[0,\frac{1}{\alpha}(N+\frac{\beta}{\alpha})\,\mathrm{tr}(S_T\Psi \Omega^D \Psi^T)])\rightarrow 0$$
\hfill$\square$
\end{corollary}
\subsection{Approximation of S-MPC with constant gains}
\label{sec:num_implementation_constant}
The solution presented in this section is characterized by a great simplicity and consists in setting $L_t=\bar{L}$ and $K_t=\bar{K}$ for all $t\geq 0$. In this case, the value of $\Sigma_{t+k}$ (and therefore of $X_{t+k}$ and $U_{t+k}$) can be directly computed for any $k>0$ by means of~\eqref{eq:variance_evolution_error}  as soon as $\Sigma_t$ is given. As a byproduct, the nonlinearity in the constraints \eqref{eq:Cantelli_ineqs} does not carry about implementation problems. Therefore, this solution has a twofold advantage: first, it is simple and requires an extremely lightweight implementation; secondly, it allows for the use of nonlinear less conservative constraint formulations. In this simplified framework, the following S-MPCc problem can be stated.\\\\
\textbf{S-MPCc problem: }at any time instant $t$ solve
    $$\min_{\bar{x}_{t},\Sigma_{t},\bar{u}_{t, \dots, t+N-1}} J$$
where $J$ is defined in \eqref{eq:JTOT}, \eqref{eq:mean_cost_function}, \eqref{eq:variance_cost_function1}, subject to
\begin{itemize}
  \item[-] the dynamics~\eqref{eq:mean_value_evolution_freeIC} , with $K_t=\bar{K}$, and
  \begin{equation}
\Sigma_{t+1}=\Phi\Sigma_{t}\Phi^T+\Psi\Omega\Psi^T
\end{equation}
  \item[-] the constraints~\eqref{eq:Cantelli_ineqs} for all $k=0,\dots,N-1$;
  \item[-] the initialization constraint \eqref{eq:reset_constraint};
    \item[-] the terminal constraints~\eqref{eq:term_constraint_mean}, \eqref{eq:term_constraint_variance}.
\end{itemize}
\hfill$\square$\\
An additional remark is due. The term $J_v$ in~\eqref{eq:variance_cost_function} does not depend only of the control and observer gain sequences $K_{t,\dots, t+N-1}$,
$L_{t,\dots, t+N-1}$, but also of the initial condition $\Sigma_t$. Therefore, it is not possible to discard it in this simplified formulation.\\
The following corollary can be derived from Theorem~\ref{thm:main}.
\begin{corollary}\label{cor:const_gains}
If, at time $t=0$, the S-MPCc problem admits a solution, the optimization problem is recursively feasible and the state and input probabilistic constraints \eqref{eq:prob_constraint_state} and \eqref{eq:prob_constraint_input} are satisfied for all $t\geq 0$. Furthermore, if there exists $\rho\in(0,1)$ such that the noise variance $\Omega$ verifies~\eqref{eq:cons_conds_on_W}, then, as $t\rightarrow +\infty$
$$\mathrm{dist}(\|\bar{x}_t\|^2+\mathrm{tr}\{\Sigma_{t|t}\},[0,\frac{1}{\alpha}(N+\frac{\beta}{\alpha})\,\mathrm{tr}(S_T\Psi \Omega \Psi^T)])\rightarrow 0$$
\hfill$\square$
\end{corollary}
%For the sake of clarity of presentation, the proof of Corollary \ref{cor:const_gains} can be found in Appendix \ref{app:proof_const_gains}.
%
%
\subsection{Boundedness of the input variables}
\label{sec:inputs}
The S-MPC scheme described in the previous sections does not guarantee the satisfaction of hard constraints on the input variables. However, input variables can be bounded in practice, and subject to
\begin{align}
Hu_t\leq \mathbf{1}
\label{eq:input_constr}
\end{align}
where $H\in\mathbb{R}^{n_H\times m}$ is a design matrix and $\mathbf{1}$ is a vector of dimension $n_H$ whose entries are equal to $1$.
Three possible approaches are proposed to account for these inequalities.\\
1) Inequalities \eqref{eq:input_constr} can be stated as additive probabilistic constraints~\eqref{eq:prob_constraint_input} with small violation probabilities $p_s^u$. This solution, although not guaranteeing satisfaction of~\eqref{eq:input_constr} with probability 1, is simple and of easy application.\\
2) If the S-MPCc scheme is used, define the gain matrix $\bar{K}$ in such a way that $A-B\bar{K}$ is asymptotically stable and, at the same time, $H\bar{K}=0$. From \eqref{eq:fb_control_law_ideal}, it follows that
$Hu_t=H\bar{u}_t+H\bar{K}(\hat{x}_t-\bar{x}_t)=H\bar{u}_t$. Therefore, to verify \eqref{eq:input_constr} it is sufficient to include in the problem formulation the deterministic constraint $H\bar{u}_t\leq \mathbf{1}$.\\
3) In an S-MPCc scheme, in case probabilistic constraint on the input variables were absent, we can replace \eqref{eq:fb_control_law_ideal} with $u_t=\bar{u}_t$ and set $H\bar{u}_t\leq \mathbf{1}$ in the S-MPC optimization problem to verify \eqref{eq:input_constr}. If we also define $\hat{u}_t=\bar{u}_t-\bar{K}(\hat{x}_t-\bar{x}_t)$ as the input to equation \eqref{eq:observer1}, the dynamics of variable $\sigma_t$ is given by \eqref{eq:error_matrix} with
$$\Phi_t=\begin{bmatrix}A-\bar{L}C&B\bar{K}\\
          \bar{L}C&A-B\bar{K}\end{bmatrix}$$
and the arguments follow similarly to those proposed in the paper. It is worth mentioning, however, that matrix $\Phi_t$ must be asymptotically stable, which requires asymptotic stability of $A$.
%A remark has been devoted, in the paper, to illustrating the three solutions described above. As future work, we commit ourselves to further explore these solutions, with special focus on the third approach.
%
%
%
%
%
%%%%%%%%%%%%%%%%%%%%%%%%%%%%%%%%%%%%%%%%%%%%
%
%
%
\section{Examples}
\label{sec:example}
In this section a comparison between the characteristics of the proposed method and the well-known robust tube-based MPC is first discussed. Then, the approximations described in Section \ref{sec:num_implementation} are discussed with reference to a numerical example.
\subsection{Simple analytic example: comparison between the probabilistic and the deterministic robust MPC}
\label{app:example_constrs}
\begin{figure}
    \centering
    \includegraphics[width=0.6\linewidth]{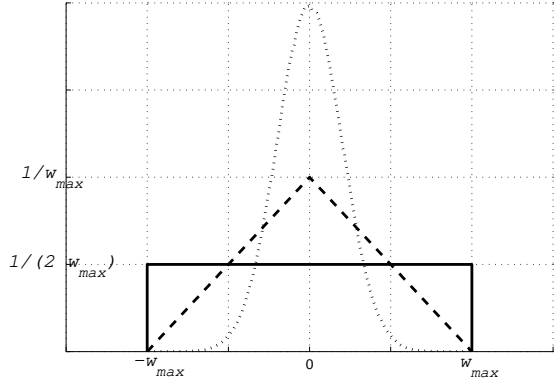}
    \caption{Distributions: uniform (case A, solid line), triangular (case B, dashed line), truncated Gaussian (case C, dotted line).}
    \label{fig:distrs}
\end{figure}

Consider the scalar system
$$x_{t+1}=a x_t+u_t+w_t$$
where $0<a<1$, $w\in[-w_{max},w_{max}]$, $w_{max}>0$, and the measurable state is constrained as follows
\begin{align}x_t\leq x_{max}\label{eq:ex_det_constr}
\end{align}
The limitations imposed by the deterministic robust MPC algorithm developed in \cite{mayne2005robust} and by the probabilistic (state-feedback) method described in this paper are now compared.
For both the algorithms, the control law $u_t=\bar{u}_t$ is considered, where $\bar{u}$ is the input of the nominal/average system $\bar{x}_{t+1}=a\bar{x}_{t}+b\bar{u}_{t}$ with state $\bar{x}$. Note that this control law is equivalent to \eqref{eq:fb_control_law_ideal}, where for simplicity it has been set $K_{t}=0$.\\
In the probabilistic approach, we allow the constraint \eqref{eq:ex_det_constr} to be violated with probability $p$, i.e.,
\begin{align}\mathbb{P}\{x\geq x_{ max}\}\leq p\label{eq:ex_prob_constr}
\end{align}
To verify \eqref{eq:ex_det_constr} and \eqref{eq:ex_prob_constr} the tightened constraint $\bar{x}_k\leq x_{max}-\Delta x$ must be fulfilled in both the approaches where, in case of \cite{mayne2005robust}, $\Delta x =\Delta x_{RPI}=\sum_{i=0}^{+\infty}a^iw_{max}=\frac{1}{1-a} w_{max}$ while, having defined $w$ as a stochastic process with zero mean and variance $W$, in the probabilistic framework $\Delta x=\Delta x_{S}(p) =\sqrt{X(1-p)/p}$, and  $X$ is the steady state variance satisfying the algebraic equation $X=a^2X+W$, i.e. $X=W/(1-a^2)$. Notably, $W$ takes different values depending upon the noise distribution.\\
It results that the deterministic tightened constraints are more conservative provided that $\Delta x_{S}(p)<\Delta x_{RPI}$, i.e.
\begin{equation}p>\frac{(1-a)^2}{b(1-a^2)+(1-a)^2}\label{eq:ex_pbound}\end{equation}
Consider now the distributions depicted in Figure \ref{fig:distrs} with $W=w_{max}^2/b$, where\\
Case A) $b=3$ for uniform distribution;\\
Case B) $b=18$ for triangular distribution;\\
Case C) $b=25$ for truncated Gaussian distribution.\\
Setting, for example, $a=0.9$, condition \eqref{eq:ex_pbound} is verified for $p>0.0172$ in case A), $p>0.0029$ in case B), and $p>0.0021$ in case C). Note that, although formally truncated, the distribution in case C) can be well approximated with a non-truncated Gaussian distribution: if this information were available, one could use $\Delta x_S(p)=\sqrt{X}\,\mathcal{N}^{-1}(1-p)$ for constraint tightening, and in this case $\Delta x_{S}(p)<\Delta x_{RPI}$ would be verified with $p>1-\mathcal{N}\left(\frac{(1-a^2)b}{(1-a)^2}\right)\simeq 0$.
\subsection{Simulation example}
\begin{figure}
    \centering
    \includegraphics[width=1\linewidth]{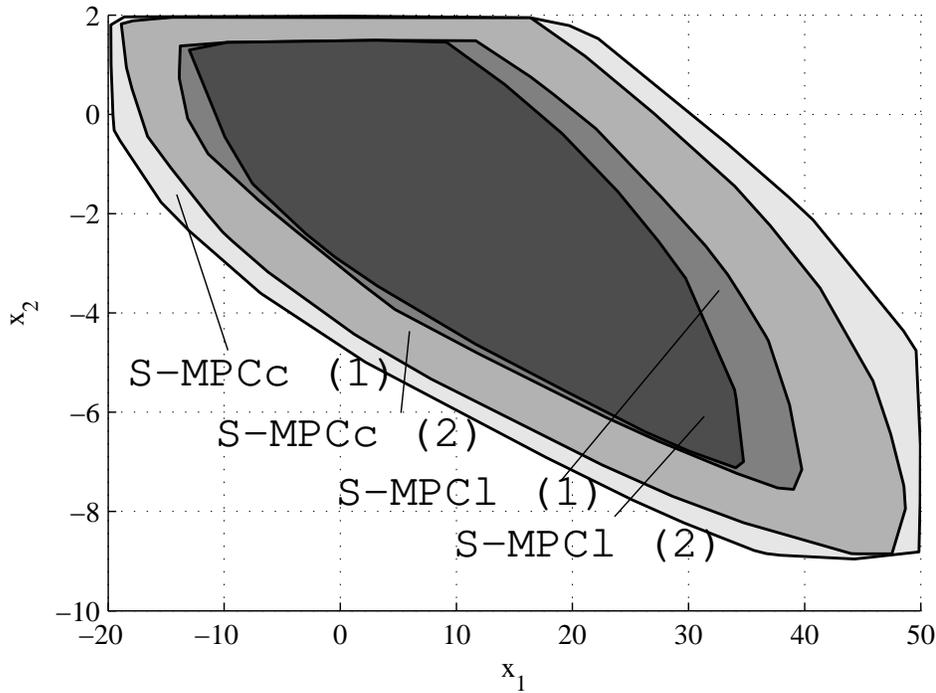}
    \caption{Plots of the feasibility sets for S-MPCc (1), S-MPCc (2), S-MPCl (1), S-MPCl (2)}
    \label{fig:sets}
\end{figure}

The example shown in this section is inspired by \cite{mayne2005robust}. We take
$$A=\begin{bmatrix}1&1\\0&1\end{bmatrix}, B=\begin{bmatrix}0.5\\1\end{bmatrix}$$
$F=I_2$, and $C=I_2$. We assume that a Gaussian noise affects the system, with $W=0.01I_2$ and $V=10^{-4}I_2$. The chance-constraints are $\mathbb{P}\{x_{2}\geq 2\}\leq 0.1$, $\mathbb{P}\{u\geq 1\}\leq 0.1$, and $\mathbb{P}\{-u\geq 1\}\leq 0.1$. In \eqref{eq:JTOT}, \eqref{eq:mean_cost_function}, and \eqref{eq:variance_cost_function} we set $Q_L=Q=I_2$, $R=0.01$, and $N=9$.\\
In Figure \ref{fig:sets} we compare the feasible sets obtained with the different methods presented in Section \ref{sec:num_implementation}, with different assumptions concerning the noise (namely S-MPCc (1), S-MPCc (2), S-MPCl (1), S-MPCl (2), where (1) denotes the case of Gaussian distribution and (2) denotes the case when the distribution is unknown). Apparently, in view of the linearization of the constraints (see the discussion after \eqref{eq:Cantelli_ineqs_linL}), the S-MPCl algorithm results more conservative than S-MPCc. On the other hand, concerning the dimension of the obtained feasibility set, in this case the use of the Chebyshev - Cantelli inequality does not carry about a dramatic performance degradation in terms of conservativeness.\\
%In Figure \ref{fig:x1x2} we show the evolution of the state variables $x_1$ and $x_2$, respectively, using the different control approaches, for 200 Montecarlo runs, starting from initial condition $(5,-1.5)$. Also, in Figure \ref{fig:u} we show the corresponding inputs. Apparently, the fact that the control and estimation gains are free variables makes the transient behaviour of the state responses in case of S-MPCl more performing, with respect to the case when S-MPCc is used. For a more detailed analysis, please see Table \ref{Table:res_sim}, where it is witnessed that the overshoot and the variance of the dynamic state response are reduced in case of S-MPCl, at the price of a more reactive input response.\\
%\begin{figure}
%    \centering
%    \includegraphics[width=.49\columnwidth]{x1}
%    \includegraphics[width=.49\columnwidth]{x2}
%    \caption{Trajectories $x_1$ (left) and $x_2(k)$ (right) for $200$ runs (light grey lines), sampled mean value (black solid line line), mean value $\pm$ sampled standard deviation (dotted dark grey lines), maximum and minimum values (dashed dark grey lines).}
%    \label{fig:x1x2}
%\end{figure}
%\begin{figure}
%    \centering
%    \includegraphics[width=.49\columnwidth]{u}
%    \caption{Trajectories $u(k)$ for $200$ runs (light grey lines), sampled mean value (black solid line line), mean value $\pm$ sampled standard deviation (dotted dark grey lines), maximum and minimum values (dashed dark grey lines).}
%    \label{fig:u}
%\end{figure}
A 200-runs Montecarlo simulation campaign has been carried out for testing the probabilistic properties of the algorithm, with initial conditions $(5,-1.5)$. The fact that the control and estimation gains are free variables makes the transient behaviour of the state responses in case of S-MPCl more performing and reduces the variance of the dynamic state response (at the price of a more reactive input response), with respect to the case when S-MPCc is used. For example, the maximum variance of $x_1(k)$ (resp. of $x_1(k)$) is about $0.33$ (resp. $0.036$) in case of S-MPCc (1) and (2), while it results about $0.25$ (resp. $0.035$) in case of S-MPCl (1) and (2). On the other hand, the maximum variance of $u(k)$ is about $0.006$ in case of S-MPCc, while it is $0.008$ in case of S-MPCl.
%
%\begin{table}[h]
%\centering
%\begin{tabular}{  l || c | c | c | c }
%  & S-MPCc (1) & S-MPCc (2) & S-MPCl (1) & S-MPCl (2) \\
%  \hline
%  Overshoot on $\mathbb{E}\{x_1(k)\}$&-0.3395  & -0.3395 & -0.2615 & -0.2649\\
%  Overshoot on $\mathbb{E}\{x_2(k)\}$&0.0608 & 0.0608 & 0.0505 & 0.0535\\
%  Overshoot on $\mathbb{E}\{u(k)\}$&-0.0125 & -0.0125 & -0.0102& -0.0106\\
%  Max variance of $x_1(k)$&0.3298 & 0.3298 & 0.2521& 0.2498\\
%  Max variance of $x_2(k)$&0.0358 &  0.0358 & 0.0355&0.0354\\
%  Max variance of $u(k)$&0.0058  &0.0058 & 0.0080&0.0080
%\end{tabular}
%\label{Table:res_sim}
%\caption{Comparison of the dynamic behaviour of the trajectories using the different approaches.}
%\end{table}
\section{Conclusions}
\label{sec:conclusions}
The main properties of the proposed probabilistic MPC algorithm lie in its simplicity and in its light-weight computational load, both in the off-line design phase and in the online implementation. This allows for the application of the S-MPC scheme to medium/large-scale problems, for general systems affected by general disturbances.\\
Future work will focus on the use of the proposed scheme in challenging control problems, such as the control of micro-grids in presence of renewable stochastic energy sources. The application of the algorithm to cope with linear time-varying systems is envisaged, while its extension to distributed implementations is currently underway.
\section*{Acknowledgements}
We are indebted with Bruno Picasso for fruitful discussions and suggestions.
\appendix
%\section{Reformulation of the chance-constraints}
%\subsection{Proof of Proposition \ref{prop:Cantelli}}
%\label{app:proofProp}
%%    For the proof of Proposition \ref{prop:Cantelli} we use the following result.
%%    \begin{lemma}
%%    \textbf{(Cantelli inequality)} Let $y$ be a scalar random variable with
%%    mean $\bar{y}$ and variance $Y$. Then for every $\mathbb{R}\ni\alpha\geq 0$ it holds that
%%    \begin{equation}\label{eq:cantelli}
%%    \mathbb{P}(y\geq \overline{y}+\alpha )\leq \frac{Y}{Y+\alpha ^{2}}
%%    \end{equation}\hfill$\blacksquare$
%%    \end{lemma}
%    Consider the constraints \eqref{eq:prob_constraint_state} and \eqref{eq:prob_constraint_input} and note that they both take the general form \eqref{eq:prob_constraint_general}. We write
%    $$\mathbb{P}\{h^T z\geq z^{ max}\}=\mathbb{P}\{h^T z\geq h^T \bar{z}+\alpha\}$$
%    where $\alpha=z^{ max}- h^T \bar{z}$. For $z^{ max}\geq h^T \bar{z}$ we resort to the Cantelli inequality, see e.g. \cite{Cantelli}, and we obtain that
%    $$\mathbb{P}\{h^T z\geq z^{ max}\}\leq \frac{h^T Z h}{h^T Z h+(z^{ max}- h^T \bar{z})^2}$$
%    Therefore, the original inequality \eqref{eq:prob_constraint_general} is satisfied by imposing
%    \begin{equation}\label{eq:state_variance_constraint}
%    \frac{h^T Z h}{h^T Z h+(z^{ max}- h^T \bar{z})^2}\leq p
%    \end{equation}
%    i.e., by requiring \eqref{eq:Cantelli_propGen} with $f(p)=\sqrt{\frac{1-p}{p}}$.
	%
\section{Proof of the main Theorem \ref{thm:main}}
\label{app:proof_Theorem}
    Recursive feasibility is first proved. Assume that, at time instant $t$, a feasible solution of S-MPC is available, i.e., $(\bar{x}_{t|t},\Sigma_{t|t})\in\Xi$ with optimal sequences $\bar{u}_{t,\dots, t+N-1|t}$,
	$K_{t,\dots, t+N-1|t}$, and $L_{t,\dots, t+N-1|t}$. We prove that at time $t+1$ a feasible solution exists, i.e., in view of Strategy 2, $(\bar{x}_{t+1|t},\Sigma_{t+1|t})\in\Xi$ with admissible sequences
	$\bar{u}^f_{t+1,\dots, t+N|t}=\{\bar{u}_{t+1|t},\dots, \bar{u}_{t+N-1|t},$ $-\bar{K}\bar{x}_{t+N|t}\}$,
	$K^f_{t+1,\dots, t+N|t}=\{K_{t+1|t},\dots,K_{t+N-1|t},\bar{K}\}$, and\break
    $L^f_{t+1,\dots, t+N|t}=\{L_{t+1|t},$ $\dots, L_{t+N-1|t},\bar{L}\}$.
	Constraint \eqref{eq:Cantelli_ineqs_state} is verified for all pairs $(\bar{x}_{t+1+k|t},X_{t+1+k|t})$, $k=0,\dots,N-2$, in view of the feasibility of S-MPC at time $t$. Furthermore, in view of \eqref{eq:term_constraint_mean}, \eqref{eq:term_constraint_variance}, \eqref{eq:Xbar_def}, and the condition \eqref{eq:linear_constraint_state_finalc},
	we have that
	$$\begin{array}{lcl}b^T\bar{x}_{t+N|t}&\leq& x^{ max}-\sqrt{b^T\bar{X}b} f(p_r^x)\\
	&\leq&
	x^{ max}-\sqrt{b^T X_{t+N|t} b} f(p_r^x)\end{array}$$
	i.e., constraint \eqref{eq:Cantelli_ineqs_state} is verified.\\
	Analogously, constraint \eqref{eq:Cantelli_ineqs_input} is verified for all pairs $(\bar{u}_{t+1+k|t},U_{t+1+k|t})$, $k=0,\dots,N-2$, in view of the feasibility of S-MPC at time $t$. Furthermore, in view of \eqref{eq:term_constraint_mean}, \eqref{eq:term_constraint_variance}, \eqref{eq:Ubar_def}, and the condition \eqref{eq:linear_constraint_input_finalc},
	we have that
	$$\begin{array}{lcl}
	-c^T\bar{K}\bar{x}_{t+N|t}&\leq& u^{ max}-\sqrt{c^T\bar{U}c} f(p_s^u)\\
	&\leq&
	u^{ max}-\sqrt{c^T{U}_{t+N|t}c}f(p_s^u)
	\end{array}$$
	i.e., constraint \eqref{eq:Cantelli_ineqs_input} is verified.\\
	In view of \eqref{eq:term_constraint_mean} and of the invariance property \eqref{eq:inv_terminal} it follows that $\bar{x}_{t+N+1|t}=(A-B\bar{K})\bar{x}_{t+N|t}\in\bar{\mathbb{X}}_F$ and, in view of \eqref{eq:term_constraint_variance}, \eqref{eq:Riccati_1}
   	$$\begin{array}{lcl}
	 \Sigma_{t+N+1|t}&=&\Phi
    \Sigma_{t+N|t}\Phi^T+\Psi\Omega\Psi^T\\
	&\preceq& \Phi\bar{\Sigma}\Phi^T+\Psi\bar{\Omega}\Psi^T=\bar{\Sigma}\end{array}$$
	hence verifying both \eqref{eq:term_constraint_mean} and \eqref{eq:term_constraint_variance} at time $t+1$.

	The proof of convergence is partially inspired by \cite{MagniRaiScatt,RaiEtAl_ECC09}. In view of the feasibility, at time $t+1$ of the possibly suboptimal solution $\bar{u}^f_{t+1,\dots, t+N|t}$,
	$K^f_{t+1,\dots, t+N|t}$, $L^f_{t+1,\dots, t+N|t}$, and ($\bar{x}_{t+1|t},\Sigma_{t+1|t}$), we have that the optimal cost function computed at time $t+1$ is
	$J^*(t+1)=J^*_m(t+1)+J_v^*(t+1)$\footnote{For brevity, we denote $J^*(x_{t},\bar{x}_{t|t-1},\Sigma_{t|t-1})$ with $J^*(t)$, $J^*_m(x_{t},\bar{x}_{t|t-1},\Sigma_{t|t-1})$ with $J^*_m$(t), and $J_v^*(x_{t},\bar{x}_{t|t-1},\Sigma_{t|t-1})$ with $J_v^*(t)$}. In view of the optimality of $J^*(t+1)$
	\begin{align}
	J^*(t+1)&\leq J_m(\bar{x}_{t+1|t},\bar{u}^f_{t+1,\dots, t+N|t})\nonumber\\
	&+ J_v(\Sigma_{t+1|t},K^f_{t+1,\dots, t+N|t},L^f_{t+1,\dots, t+N|t})\label{eq:J<J+J1}\end{align}
	Note that
	 \begin{align}&J_m(\bar{x}_{t+1|t},\bar{u}^f_{t+1,\dots, t+N|t})=\nonumber\\
	&J_m(\bar{x}_{t|t},\bar{u}_{t,\dots, t+N-1|t})-\|\bar{x}_{t|t}\|_Q^2-\|\bar{u}_{t|t}\|_R^2+\nonumber\\
	 &\|\bar{x}_{t+N|t}\|_Q^2+\|\bar{K}\bar{x}_{t+N|t}\|_R^2-\|\bar{x}_{t+N|t}\|_S^2+\nonumber\\
	 &\|(A-B\bar{K})\bar{x}_{t+N|t}\|_{S}^2\label{eq:J<J+J2}
	\end{align}
In view of \eqref{eq:Lyap_S}	 \begin{align}&\|\bar{x}_{t+N|t}\|_Q^2+\|\bar{K}\bar{x}_{t+N|t}\|_R^2-\|\bar{x}_{t+N|t}\|_S^2+\nonumber\\
	 &\|(A-B\bar{K})\bar{x}_{t+N|t}\|_{S}^2\leq0\label{eq:J<J+J3}
	\end{align}
	Furthermore, note that
	\begin{align}J_m(\bar{x}_{t|t},\bar{u}_{t,\dots, t+N-1|t})=J_m^*(t)\label{eq:J<J+J4}
	\end{align}
	Now consider $J_v$ in \eqref{eq:variance_cost_function} and note that
	\begin{align}&J_v(X_{t+1|t},K^f_{t+1,\dots, t+N|t},L^f_{t+1,\dots, t+N|t})\nonumber\\
	&=J_v(X_{t|t},K_{t,\dots, t+N-1|t},L_{t,\dots, t+N-1|t})\nonumber\\
&-\mathrm{tr}\{\begin{bmatrix}Q_L&0\\0&Q+K_{t|t}^TRK_{t|t}
\end{bmatrix}\Sigma_{t|t}\}+\mathrm{tr}\{Q_T\Sigma_{t+N|t}\label{eq:J<J+J5}\\ &+S_T\Phi\Sigma_{t+N|t}\Phi^T+S_T\Psi\Omega\Psi^T-S_T\Sigma_{t+N|t}\}\nonumber
	\end{align}
	Recalling the properties of the trace and \eqref{eq:Lyap_S}, one has:
	 \small\begin{align}&\mathrm{tr}\{Q_T\Sigma_{t+N|t}+S_T\Phi\Sigma_{t+N|t}\Phi^T-S_T\Sigma_{t+N|t}\}=\nonumber\\
	&\mathrm{tr}\{(Q_T+\Phi^TS_T\Phi-S_T)\Sigma_{t+N|t}\}\leq 0\label{eq:J<J+J6}
	\end{align}\normalsize
	From \eqref{eq:J<J+J1}-\eqref{eq:J<J+J6} we obtain
	 \small\begin{equation}\begin{array}{ll}J^*(t+1)&\leq J^*(t)-(\|\bar{x}_{t|t}\|^2_Q+\|\bar{u}_{t|t}\|^2_R)\\
&-\mathrm{tr}\{\begin{bmatrix}Q_L&0\\0&Q+K_{t|t}^TRK_{t|t}
\end{bmatrix}\Sigma_{t|t}\}+\mathrm{tr}(S_T\Psi\Omega\Psi^T)\end{array}
	 \label{eq:ineq_J}
	\end{equation}\normalsize
	Furthermore, from the definition of $J^*(t)$ we also have that \begin{align}J^*(t)&\geq\|\bar{x}_{t|t}\|^2_Q+\|\bar{u}_{t|t}\|^2_R\nonumber\\
&+\mathrm{tr}\left\{\begin{bmatrix}Q_L&0\\0&Q+K_{t|t}^TRK_{t|t}
\end{bmatrix}\Sigma_{t|t}\right\}\label{eq:ineq_Jgeq}
	\end{align}
	Now, denote $\Omega_F=\{(\bar{x},\Sigma):\bar{x}\in\bar{\mathbb{X}}_F,\Sigma\preceq\bar{\Sigma}\}$. Assuming that $(\bar{x}_{t|t},\Sigma_{t|t})\in\Omega_F$ we have that $J^*(t)\leq J^{aux}_m(t)+J_v^{aux}(t)$, where
	$$\begin{array}{ll}
	J_m^{aux}(t)&= \sum_{k=0}^{N-1}\|(A-B\bar{K})^{k}\bar{x}_{t|t}\|_Q^2\\	 &+\|\bar{K}(A-B\bar{K})^{k}\bar{x}_{t|t}\|_R^2+\|(A-B\bar{K})^{N}\bar{x}_{t|t}\|_S^2
	\end{array}$$
	since $\{-\bar{K}\bar{x}_{t|t},\dots,-\bar{K}(A-B\bar{K})^{N-1}\bar{x}_{t|t}\}$ is a feasible input sequence. Therefore, from \eqref{eq:Lyap_S},
	\begin{equation}
	J_m^{aux}(t)\leq \|\bar{x}_{t|t}\|_S^2\label{eq:ineq_Jmleq}
	\end{equation}
	Similarly, and recalling the properties of the trace and \eqref{eq:Lyap_S}, we obtain that $J_v^{aux}(t)$ is equal to
	\small$$\begin{array}{l} \sum_{k=0}^{N-1}\mathrm{tr}\{Q_T[\Phi^{k}\Sigma_{t|t}\Phi^{T(k)}+\sum_{i=0}^{k-1}\Phi^{i}\Psi\Omega\Psi^T\Phi^{T(i)}]\}\\	 +\mathrm{tr}\{S_T[\Phi^{N}\Sigma_{t|t}\Phi^{T(N)}+\sum_{i=0}^{N-1}\Phi^{i}\Psi\Omega\Psi^T\Phi^{T(i)}]\}\\
	= \mathrm{tr}\{[\sum_{k=0}^{N-1} \Phi^{T(k)}Q_T\Phi^{k}+\Phi^{T(N)}S_T\Phi^{N}]\Sigma_{t|t}\}\\
	 +\mathrm{tr}\{[\sum_{k=1}^{N-1}\sum_{i=0}^{k-1}\Phi^{T(i)}Q_T\Phi^{i}
+\sum_{i=0}^{N-1}\Phi^{T(i)}S_T\Phi^{i}]\Psi\Omega\Psi^T\}\\
	 \leq
\mathrm{tr}\{S_T\Sigma_{t|t}\}+\mathrm{tr}\{[S_T+\sum_{k=1}^{N-1}(\Phi^{T(k)}S_T\Phi^{k})\\
	 +\sum_{i=1}^{N-1}\Phi^{T(i)}Q_T\Phi^{i}]\Psi\Omega\Psi^T\}\\
	\leq \mathrm{tr}\{S_T\Sigma_{t|t}\}+\mathrm{tr}\{[S_T+\sum_{i=1}^{N-1}S_T]\Psi\Omega\Psi^T\}
	\end{array}$$\normalsize
	Therefore
	\begin{equation}
	J_v^{aux}(t)\leq \mathrm{tr}\{S_T \Sigma_{t|t}\}+Ntr\{S_T \Psi\Omega\Psi^T\}
	\label{eq:ineq_Jvleq}
	\end{equation}
	Combining \eqref{eq:ineq_Jmleq} and \eqref{eq:ineq_Jvleq} we obtain that, for all $(\bar{x}_{t|t},\Sigma_{t|t})\in\Omega_F$
	\begin{equation}J^*(t)\leq
	\|\bar{x}_{t|t}\|^2_S+\mathrm{tr}\{S_T \Sigma_{t|t}\}+N\,\mathrm{tr}\{S_T \Psi\Omega\Psi^T\}\label{eq:ineq_Jleq}
	\end{equation}
	%Remark that assumptions \eqref{eq:ineq_J}, \eqref{eq:ineq_Jgeq}, and \eqref{eq:ineq_Jleq} are similar to the ones needed in the framework of input-to-state stability.
	From \eqref{eq:ineq_J}, \eqref{eq:ineq_Jgeq} and \eqref{eq:ineq_Jleq} it is possible to derive robust stability-related results.\\
Before to proceed, recall that $\mathrm{tr}\{S_T \Sigma_{t|t}\}=\mathrm{tr}\{S_T^{\frac{1}{2}T}\Sigma_{t|t}S_T^{\frac{1}{2}}\}$ where $S_T^{\frac{1}{2}}$ is a matrix that verifies $S_T^{\frac{1}{2}T}S_T^{\frac{1}{2}}=S_T$. Therefore $\mathrm{tr}\{S_T \Sigma_{t|t}\}=\mathrm{tr}\{S_T^{\frac{1}{2}T}\Sigma_{t|t}S_T^{\frac{1}{2}}\}=\|\Sigma_{t|t}^{\frac{1}{2}}S_T^{\frac{1}{2}}\|^2_F$. On the other hand, denoting
$Q_{T|t}=\mathrm{diag}(Q_L,Q+{K}_{t|t}^TR{K}_{t|t})$, it follows that $\mathrm{tr}\{Q_{T|t} \Sigma_{t|t}\}=\|\Sigma_{t|t}^{\frac{1}{2}}Q_{T|t}^{\frac{1}{2}}\|^2_F$.
Recall that the sub-moltiplicativity property holds also for the Frobenius norm, implying that $\|AB\|_F\leq \|A\|_F\|B\|_F$ and that $\|AB\|_F\geq (\|A^{-1}\|_F)^{-1}\|B\|_F$. In view of this, it follows that $\mathrm{tr}\{S_T \Sigma_{t|t}\}\leq\|\Sigma_{t|t}^{\frac{1}{2}}\|^2_F\|S_T^{\frac{1}{2}}\|^2_F=\mathrm{tr}\{S_T\}\mathrm{tr}\{\Sigma_{t|t}\}$ and that, also considering the matrix inversion Lemma,
$\mathrm{tr}\{Q_{T|t} \Sigma_{t|t}\}\geq
(\|Q_{T|t}^{-\frac{1}{2}}\|^2_F)^{-1}\|\Sigma_{t|t}^{\frac{1}{2}}\|^2_F\geq
\mathrm{tr}\{(\mathrm{diag}(Q_L,Q))^{-1}\}^{-1}\mathrm{tr}\{\Sigma_{t|t}\}=\mathrm{tr}\{Q^{-1}+Q_L^{-1}\}^{-1}\mathrm{tr}\{\Sigma_{t|t}\}$.\\
Define $V(\bar{x}_{t|t},\Sigma_{t|t})=\|\bar{x}_{t|t}\|^2+\mathrm{tr}\{\Sigma_{t|t}\}$ and $\omega=\mathrm{tr}\{S_T\Psi\Omega\Psi^T\}$. In view of this, we can reformulate \eqref{eq:ineq_J}, \eqref{eq:ineq_Jgeq} and \eqref{eq:ineq_Jleq} as follows.
\begin{subequations}
    \begin{align}
    J^*(t+1)&\leq J^*(t)-\alpha V(\bar{x}_{t|t},\Sigma_{t|t})+\omega\label{eq:ineq_J_b}\\
    J^*(t)&\geq \alpha V(\bar{x}_{t|t},\Sigma_{t|t})\label{eq:ineq_Jgeq_b}\\
    J^*(t)&\leq \beta V(\bar{x}_{t|t},\Sigma_{t|t})+N\omega\label{eq:ineq_Jleq_b}
    \end{align}
    \label{eq:ineqs_J_b}
\end{subequations}
	If $(\bar{x}_{t|t},\Sigma_{t|t})\in\Omega_F$ then, in view of \eqref{eq:ineq_Jleq_b}, \eqref{eq:ineq_J_b}
	\begin{align}
	J^*(t+1)\leq& J^*(t)(1-\frac{\alpha}{\beta})+(\frac{\alpha}{\beta}N+1)\omega\label{eq:ineq_JJ}
	\end{align}
	Let $\eta\in(\rho,1)$ and denote $b=\frac{1}{\eta}(N+\frac{\beta}{\alpha})$. In view of \eqref{eq:ineq_Jgeq_b}, if $J^*(t)\leq b\, \omega$
	$$V(\bar{x}_{t|t},\Sigma_{t|t})\leq \frac{b}{\alpha}\omega$$
	This, considering \eqref{eq:cons_conds_on_W}, implies that
	\begin{subequations}
	\begin{align}
	 \|\bar{x}_{t|t}\|^2&\leq \frac{\rho}{\eta}\bar{\sigma}^2\label{eq:cons_conds_on_W_x2}\\	 \mathrm{tr}(\Sigma_{t|t})&\frac{\rho}{\eta}\lambda_{min}(\bar{\Sigma})\label{eq:cons_conds_on_W_X2}
	\end{align}
	\label{eq:cons_conds_on_W2}
	\end{subequations}
	In view of \eqref{eq:cons_conds_on_W_x2}, then $\bar{x}_{t|t}\in\bar{\mathbb{X}}_F$. Furthermore, \eqref{eq:cons_conds_on_W_X2} implies that $\lambda_{max}(\Sigma_{t|t})<\lambda_{min}(\bar{\Sigma})$, which in turn implies that $\Sigma_{t|t}<\bar{\Sigma}$.
	Therefore, recalling \eqref{eq:ineq_JJ}, if $J^*(t)\leq b\, \omega$, then $J^*(t+1)\leq b \,\omega$
	and the positive invariance of the set
	\begin{equation}
	D=\{(\bar{x},\Sigma):J^*(t)\leq b\, \omega\}
	\end{equation}
	is guaranteed.\\
	If $(\bar{x}_{t|t},\Sigma_{t|t})\in \Omega_F\backslash D$, it holds that
	\begin{equation}J^*(t)> b\, \omega\label{eq:J>...}\end{equation}
	which, in view of \eqref{eq:ineq_Jleq_b}, implies that
	\begin{equation}V(\bar{x}_{t|t},\Sigma_{t|t})>\frac{1}{\alpha}\omega
    \label{eq:J-J<0}\end{equation}
    Since $(\bar{x}_{t|t},\Sigma_{t|t})\in \Omega_F\backslash D$, recalling \eqref{eq:ineq_JJ}, \eqref{eq:J-J<0}, and \eqref{eq:ineq_Jgeq_b}, there exists $\bar{c}_1>0$ (function of $\eta$) such that
	\begin{align}
	&J^*(t+1)-J^*(t)\leq -(1-\eta)\frac{\alpha}{\beta}J^*(t)\nonumber\\
	&\leq -(1-\eta)\frac{\alpha^2}{\beta}V(\bar{x}_{t|t},\Sigma_{t|t})\leq -\bar{c}_1\label{eq:ineq_JJ2}
	\end{align}
On the other hand, for all $x_t$ with $(\bar{x}_{t|t},\Sigma_{t|t})\in \Xi\backslash\Omega_F$, there exists constant $\bar{c}_2>0$ such that
there exists $x_{\Omega}$ with $(\bar{x}_{\Omega},\Sigma_{\Omega})\in \Omega_F\backslash D$ such that
	$-\alpha V(\bar{x}_{t|t},\Sigma_{t|t})\leq -\alpha V(\bar{x}_{\Omega},\Sigma_{\Omega})-\bar{c}_2$. This, in view of \eqref{eq:ineq_J_b} and \eqref{eq:ineq_JJ2}, implies that
	 \begin{align}J^*(t+1)-J^*(t)<-\bar{c}_2\label{eq:ineq_JJ1}\end{align}
In view of \eqref{eq:ineq_JJ2}-\eqref{eq:ineq_JJ1}, for all $x_t$ with $(\bar{x}_{t|t},\Sigma_{t|t})\in \Xi\backslash D$ there exists $\bar{c}$ (function of $\eta$)
	 \begin{align}J^*(t+1)-J^*(t)<-\bar{c}\label{eq:ineq_JJ3}\end{align}
This implies that, for each $\eta\in(\rho,1)$, there exists $T>0$ such that $x_{t+T}$ is such that $(\bar{x}_{t+T|t+T},\Sigma_{t+T|t+T})\in D$, i.e., that $\alpha V(\bar{x}_{t+k|t+k},\Sigma_{t+k|t+k})\leq b\omega$ for all $k\geq T$. This, for $\eta\rightarrow 1$, implies \eqref{eq:thm_stat}.
	%
	%
%\section{The S-MPCsf problem}
%\label{app:sf}
%%
%%
%\subsection{LMI reformulation of the update of $X_{k}$}
%Consider the update of $X_{i}$ in \eqref{eq:variance_evolution_S}.
%First we rewrite it as an inequality, i.e.,
%%	
%\begin{align}
%	X_{k+1}-
%	(A-BK_k)X_k(A-BK_k)^T-FWF^T\geq 0\label{eq:LMIS22}
%\end{align}
%%
%%
%By defining $\Xi_k=K_kX_k$ the inequality \eqref{eq:LMIS22} can be recast as an LMI having, as decision variables, $\Xi_k$, $X_{k+1}$, and $X_k$. In view of the Schur complement Lemma (see, e.g., \cite{MagniPalaScattoInt}) we obtain
%%	
%\begin{align}\label{LMI:sigma22D}
%	\begin{bmatrix}
%	X_{k+1}&
%	\begin{bmatrix}
%	(AX_k-B\Xi_k)&F
%	\end{bmatrix}\\
%	\begin{bmatrix}
%	(AX_k-B\Xi_k)^T\\F^T
%	\end{bmatrix}&
%	\begin{bmatrix}
%	X_k&0\\0&W^{-1}
%	\end{bmatrix}
%	\end{bmatrix}\geq 0
%\end{align}
%
%
%	
%\subsection{LMI reformulation of the constraints}
%
%While the constraint \eqref{eq:linear_constraint_state} is a linear inequality (and therefore it does not need to be reformulated as a suitable LMI), the inequality \eqref{eq:linear_constraint_input} needs special attention. In \eqref{eq:linear_constraint_input}, $U_k = K_{k}X_k K_{k}^T=\Xi_{k}(X_k)^{-1}\Xi_{k}^T$ can be recast as an LMI as follows:
%	\begin{align}\label{LMI:Uk}
%	 \begin{bmatrix}
%	  {U}_k&\Xi_k\\
%	  \Xi_k^T&X_k
%	 \end{bmatrix}\geq 0
%	\end{align}
%
%
%
\section{The S-MPCl problem}
\label{app:LMI}
\subsection{Proof of Lemma \ref{lemma:bound_var_main}}
\textbf{Part I.}\\
The following result is used.
	\begin{lemma}
\label{lemma:bound_var}
	Given a positive semi-definite, symmetric matrix $M$, then
	\begin{equation*}
	M=\begin{bmatrix}
	M_{11}&M_{12}\\M_{12}^T&M_{22}
	\end{bmatrix}\preceq
	\begin{bmatrix}
	2M_{11}&0\\0&2M_{22}
	\end{bmatrix}
	\end{equation*}\hfill$\square$
	\end{lemma}
\textbf{Proof of Lemma~\ref{lemma:bound_var}}
Since $M\succeq 0$, then
$$\begin{bmatrix}-x_1^T&x_2^T\end{bmatrix}M \begin{bmatrix}-x_1\\x_2\end{bmatrix}=x_1^T M_{11}x_1+x_2^T M_{22}x_2-x_1^T M_{12}x_2-x_2^T M_{12}^Tx_1 \succeq 0$$
for all $x_1,x_2$ such that $\begin{bmatrix}x_1^T&x_2^T\end{bmatrix}\neq 0$. From this, we obtain that
$$\begin{array}{lcl}\begin{bmatrix}x_1^T&x_2^T\end{bmatrix}M \begin{bmatrix}x_1\\x_2\end{bmatrix}&=&x_1^T M_{11}x_1+x_2^T M_{22}x_2+x_1^T M_{12}x_2+x_2^T M_{12}^Tx_1\\
 &\leq& 2x_1^T M_{11}x_1+2x_2^T M_{22}x_2= \begin{bmatrix}x_1^T&x_2^T\end{bmatrix}\begin{bmatrix}2M_{11}&0\\0&2M_{22}\end{bmatrix} \begin{bmatrix}x_1\\x_2\end{bmatrix}\end{array}$$
for all $x_1,x_2$ such that $\begin{bmatrix}x_1^T&x_2^T\end{bmatrix}\neq 0$. This concludes the proof of Lemma \ref{lemma:bound_var}.
\hfill$\square$\\\\
Consider now matrix $\Sigma_t$ and its block-decomposition
$$\Sigma_t=\begin{bmatrix}\Sigma_{11,t}&\Sigma_{12,t}\\\Sigma_{12,t}^T&\Sigma_{22,t}\end{bmatrix}$$
where $\Sigma_{ij,t}\in\mathbb{R}^{n\times n}$ for all $i,j=1,2$.
A bound for the time evolution of the covariance matrix $\Sigma_{t}$ is computed, iteratively, considering that	
\begin{align}\label{eq:variance_evolution_bound}
	 \Sigma_{t+1}&\preceq\Phi_t
	 \Sigma_{t}^D
	 \Phi_t^T+
	 \Psi_t
	 \Omega
	 \Psi_t^T
\end{align}
If we define $\Sigma^D_{t+1}=\mathrm{diag}(\Sigma_{11,t+1}^D,\Sigma_{22,t+1}^D)$, where
\begin{align}
	 \Sigma_{11,t+1}^D&=2(A-L_{t}C)\Sigma_{11,t}^D(A-L_tC)^T+\nonumber\\
		&FWF^T+2L_tVL_t^T\\
	 \Sigma_{22,t+1}^D&=2(A-BK_t)\Sigma_{22,t}^D(A-BK_t)^T+\nonumber\\
		&2L_tC\Sigma_{11,t}^DC^TL_t^T+2L_t V L_t^T
\end{align}
then we obtain that  $\Sigma^D_{t+1}\succeq \Sigma_{t+1}$, in view of Lemma~\ref{lemma:bound_var}.
Defining $A^D=\sqrt{2}A,B^D=\sqrt{2}B,C^D=\sqrt{2}C$ and $V^D=2V$, the latter corresponds with~\eqref{eq:SigmaDupdate}.

\textbf{Part IIa. LMI reformulation of the update of $\Sigma_{11,k}^D$}.\\
We rewrite constraint \eqref{eq:sigmaD1LMI} as
\begin{equation*}
	\begin{array}{l}
	\Sigma_{11,k+1}^D-\begin{bmatrix}
	(A^D-L_kC^D)&FW&L_kV^D
	\end{bmatrix}\begin{bmatrix}
	\Sigma_{11,k}^D&0&0\\0&W^{-1}&0\\0&0&(V^D)^{-1}
	\end{bmatrix}\begin{bmatrix}
	(A^D-L_kC^D)^T\\(FW)^T\\(L_kV^D)^T
	\end{bmatrix}\succeq 0
	\end{array}
\end{equation*}
Resorting to the Schur complement it is possible to derive the following equivalent form
\begin{equation*}
	\begin{array}{l}
	\begin{bmatrix}
	(\Sigma_{11,k}^D)^{-1}&0&0\\0&W&0\\0&0&V^D
	\end{bmatrix}-\begin{bmatrix}
	(A^D-L_kC^D)^T\\(FW)^T\\(L_kV^D)^T
	 \end{bmatrix}(\Sigma_{11,k+1}^{D})^{-1}\begin{bmatrix}
	(A^D-L_kC^D)&FW&L_kV^D
	\end{bmatrix}\succeq 0
	\end{array}
\end{equation*}
To obtain a linear inequality from the previous expression we define \begin{align}Z_k=(\Sigma_{11,k+1}^{D})^{-1}L_k\label{eq:Z_def}\end{align}
and $\tilde{\Sigma}_{11,i}^D=(\Sigma_{11,i}^{D})^{-1}$, i.e.,
\begin{equation*}
	\begin{array}{l}
	\begin{bmatrix}
	\tilde{\Sigma}_{11,k}^D&0&0\\0&W&0\\0&0&V^D
	\end{bmatrix}-\begin{bmatrix}
	(\tilde{\Sigma}_{11,k+1}^D A^D-Z_kC^D)^T\\(\tilde{\Sigma}_{11,k+1}^D F W)^T\\(Z_kV^D)^T
	 \end{bmatrix}\times\\(\tilde{\Sigma}_{11,k+1}^D)^{-1}
	\begin{bmatrix}
	(\tilde{\Sigma}_{11,k+1}^D A^D-Z_kC^D)&\tilde{\Sigma}_{11,k+1}^D F W & Z_kV^D
	\end{bmatrix}\succeq 0
	\end{array}
\end{equation*}
that can be written as a compact LMI as follows
\begin{align}\label{LMI:sigma11D}
	\begin{bmatrix}	
	\begin{bmatrix}
	\tilde{\Sigma}_{11,k}^D&0&0\\0&W&0\\0&0&V^D
	\end{bmatrix}&
	\begin{bmatrix}
	(\tilde{\Sigma}_{11,k+1}^D A^D-Z_kC^D)^T\\(\tilde{\Sigma}_{11,k+1}^D F W)^T\\(Z_kV^D)^T
	\end{bmatrix}\\
	\begin{bmatrix}
	(\tilde{\Sigma}_{11,k+1}^D A^D-Z_kC^D)&\tilde{\Sigma}_{11,k+1}^D F W & Z_kV^D
	\end{bmatrix}&
	\tilde{\Sigma}_{11,k+1}^{D}
	\end{bmatrix}\succeq 0
\end{align}
which is indeed linear with respect to the variables $\tilde{\Sigma}_{11,k}^D$, $\tilde{\Sigma}_{11,k+1}^D$, and $Z_k$. Notice that, however, in the constraints \eqref{eq:Cantelli_ineqs_linL} and in the cost function \eqref{eq:variance_cost_functionL}, the term ${\Sigma}_{11,i}^D$ appears, rather than its inverse $\tilde{\Sigma}_{11,i}^D$. To solve this issue, we define matrix $\Delta_k$ as an upper bound to ${\Sigma}_{11,k}^D$ (i.e., $\Delta_{k}\succeq\Sigma_{11,k}^D$), which can be recovered from $\tilde{\Sigma}_{11,k+1}^D$ through the following linear inequality
\begin{align}\label{LMI:Sk}
	\begin{bmatrix}	
	\Delta_{k}&I\\I&\tilde{\Sigma}_{11,k}^{D}
	\end{bmatrix}\succeq 0
\end{align}
Then, one should replace $\Sigma_{11,k}^D$ with $\Delta_{k}$ in \eqref{eq:Cantelli_ineqs_linL} and \eqref{eq:variance_cost_functionL}.\\\\
\textbf{Part IIb. LMI reformulation of the update of $\Sigma_{22,k}^D$}.\\
Consider now the inequality \eqref{eq:sigmaD2LMI}, i.e.,
\begin{align}
	\Sigma_{22,k+1}^D-
	 (A^D-B^DK_k)\Sigma_{22,k}^D(A^D-B^DK_k)^T-L_k(C^D\Sigma_{11,k}^DC^{D\, T}+V^D)L_k^T\succeq 0\label{eq:LMIS22}
\end{align}
Recalling \eqref{eq:Z_def}, \eqref{eq:LMIS22} can be rewritten as
\begin{align}
	\Sigma_{22,k+1}^D- (A^D-B^DK_k)\Sigma_{22,k}^D(A^D-B^DK_k)^T-\Sigma^D_{11,k+1}M_k\Sigma^D_{11,k+1}\succeq 0\label{eq:LMIS22_b}
\end{align}
where
\begin{align}
	 M_k&=Z_k(C^D\Sigma_{11,k}^DC^{D\, T}+V^D)Z_k^T\label{eq:Mak_definition0}
\end{align}	
By defining $\Xi_k=K_k\Sigma_{22,k}^D$, and using the matrix $\Delta_{k+1}$ in place of $\Sigma^D_{11,k+1}$, the inequality \eqref{eq:LMIS22} can be recast as a suitable LMI. In fact, in view of the Schur complement Lemma we obtain
\begin{align}\label{LMI:sigma22D}
	\begin{bmatrix}
	\Sigma_{22,k+1}^D&
	\begin{bmatrix}
	(A^D\Sigma_{22,k}^D-B^D\Xi_k)&\Delta_{k+1}
	\end{bmatrix}\\
	\begin{bmatrix}
	(A^D\Sigma_{22,k}^D-B^D\Xi_k)^T\\\Delta_{k+1}
	\end{bmatrix}&
	\begin{bmatrix}
	\Sigma_{22,k}^D&0\\0&\tilde{M}_k
	\end{bmatrix}
	\end{bmatrix}\succeq 0
\end{align}
where
\begin{align}
	 \tilde{M}_k&=M_k^{-1}\label{eq:Mak_definition1}
\end{align}	
The equation \eqref{eq:Mak_definition0} can be recast as the following inequality:
\begin{align}
        {M}_k&\succeq Z_k(C^D\Sigma_{11,k}^DC^{D\, T}+V^D)Z_k^T \label{eq:Mb_in}
\end{align}
which can be reformulated as
\begin{align}
        \begin{bmatrix}{M}_k& \begin{bmatrix}Z_kV^D & Z_kC^D\end{bmatrix}\\
         \begin{bmatrix}(Z_kV^D)^T \\ (Z_kC^D)^T\end{bmatrix}&
         \begin{bmatrix}V^D &0 \\0& \tilde{\Sigma}_{11,k}^D\end{bmatrix}
         \end{bmatrix}\succeq 0 \label{eq:Mb_in2}
\end{align}
Finally, concerning the equality \eqref{eq:Mak_definition1}, it can be solved using the approach proposed in \cite{ConeComp97}. Indeed, we solve the following LMI
\begin{align}
        \begin{bmatrix}{M}_k& I\\
        I& \tilde{M}_k
         \end{bmatrix}\succeq 0 \label{eq:Cone_LMI}
\end{align}
and, at the same time, we minimize the additional cost function
\begin{align}
        \mathrm{tr}\{{M}_k \tilde{M}_k\}\label{eq:Cone_cost}
\end{align}
The problem \eqref{eq:Cone_LMI}-\eqref{eq:Cone_cost} can be managed using the recursive cone complementarity linearization algorithm proposed in \cite{ConeComp97} with a suitable initialization.
\subsection{LMI reformulation of the constraints}
While the constraint \eqref{eq:linear_constraint_state} is a linear inequality (and therefore it does not need to be further reformulated), the inequality \eqref{eq:linear_constraint_input} needs special attention. As already remarked, in \eqref{eq:linear_constraint_input}, $U_k$ must be replaced by $\bar{U}_k$. In turn, the equality $\bar{U}_k =K_{k}\Sigma_{22,k}^DK_{k}^T=\Xi_{k}(\Sigma_{22,k}^D)^{-1}\Xi_{k}^T$ must be recast as an LMI as follows:
	\begin{align}\label{LMI:Uk}
	 \begin{bmatrix}
	  \bar{U}_k&\Xi_k\\
	  \Xi_k^T&\Sigma_{22,k}^D
	 \end{bmatrix}\succeq 0
	\end{align}

\subsection{Proof of Corollary \ref{cor:LMI_soluz}}
\label{app:proofL}
Concerning the proof of recursive feasibility, assume that, at time instant $t$, a feasible solution of S-MPCl is available, i.e., $(\bar{x}_{t|t},\Sigma^D_{11,t|t},\Sigma_{22,t|t}^D)$ with optimal sequences $\bar{u}_{t,\dots, t+N-1|t}$,
	$K_{t,\dots, t+N-1|t}$, and $L_{t,\dots, t+N-1|t}$. At time $t+1$ $(\bar{x}_{t+1|t},\Sigma^D_{11,t+1|t},\Sigma_{22,t+1|t}^D)$ is feasible, with admissible sequences
	$\bar{u}^f_{t+1,\dots, t+N|t}=\{\bar{u}_{t+1|t},\dots, \bar{u}_{t+N-1|t},$ $-\bar{K}\bar{x}_{t+N|t}\}$,
	$K^f_{t+1,\dots, t+N|t}=\{K_{t+1|t},\dots,K_{t+N-1|t},\bar{K}\}$, and $L^f_{t+1,\dots, t+N|t}=\{L_{t+1|t},$ $\dots, L_{t+N-1|t},\bar{L}\}$. This can be proved similarly to Section~\ref{app:proof_Theorem}.\\
Indeed, constraint \eqref{eq:linear_constraint_stateL} is verified for all pairs $(\bar{x}_{t+1+k|t},\Sigma^D_{11,t+k+1|t}+\Sigma_{22,t+k+1|t}^D)$, $k=0,\dots,N-2$, in view of the feasibility of S-MPCl at time $t$. Furthermore, in view of \eqref{eq:term_constraint_mean}, \eqref{eq:term_constraints_varianceL}, \eqref{eq:Riccati_1L}, \eqref{eq:linear_constraint_input_finalL}, \eqref{eq:linear_constraint_stateL} is verified also for $k=N-1$. Similarly, \eqref{eq:linear_constraint_inputL} is verified also for $k=N-1$.
	In view of \eqref{eq:term_constraints_varianceL}
   	$$\begin{array}{lcl} \Sigma_{11,t+N+1|t}^D&=&(A^D-\bar{L}C^D)\Sigma_{11,t+N|t}(A^D-\bar{L}C^D)^T
    +FWF^T+\bar{L}V^D\bar{L}^T\\
	&\preceq& (A^D-\bar{L}C^D)\bar{\Sigma}_{11}(A^D-\bar{L}C^D)^T
    +F\bar{W}F^T+\bar{L}\bar{V}\bar{L}^T=\bar{\Sigma}_{11}\\
    \Sigma_{22,t+N+1|t}^D&=&(A^D-B^D\bar{K})\Sigma_{22,t+N|t}(A^D-B^D\bar{K})^T
    +\bar{L}V^D\bar{L}^T+\bar{L}C^D \Sigma_{11,t+N|t}(C^D)^T\bar{L}^T\\
	&\preceq& (A^D-B^D\bar{K})\bar{\Sigma}_{22}(A^D-B^D\bar{K})^T
    +\bar{L}\bar{V}\bar{L}^T+\bar{L}C^D \bar{\Sigma}_{11}(C^D)^T\bar{L}^T=\bar{\Sigma}_{22}
    \end{array}$$
	hence verifying both \eqref{eq:term_constraint_mean} and \eqref{eq:term_constraints_varianceL} at time $t+1$.\\\\
Concerning the proof of convergence, in view of the feasibility, at time $t+1$ of the possibly suboptimal solution $\bar{u}^f_{t+1,\dots, t+N|t}$,
$K^f_{t+1,\dots, t+N|t}$, $L^f_{t+1,\dots, t+N|t}$, and $(\bar{x}_{t+1|t},\Sigma^D_{11,t+1|t},\Sigma_{22,t+1|t}^D)$, we have that the optimal cost function computed at time $t+1$ is
$J^*(t+1)=J^*_m(t+1)+J_v^*(t+1)$. In view of the optimality of $J^*(t+1)$ \eqref{eq:J<J+J1} holds.
From \eqref{eq:J<J+J2} and \eqref{eq:Lyap_S_L}, equation \eqref{eq:J<J+J3} still holds, as well as \eqref{eq:J<J+J4}.\\
Considering \eqref{eq:Lyap_S_L}, we obtain
\small\begin{equation}\begin{array}{ll}J^*(t+1)&\leq J^*(t)-(\|\bar{x}_{t|t}\|^2_Q+\|\bar{u}_{t|t}\|^2_R)\\
&-\mathrm{tr}\{\begin{bmatrix}Q_L&0\\0&Q+K_{t|t}^TRK_{t|t}
\end{bmatrix}\Sigma^D_{t|t}\}+\mathrm{tr}(S_T\Psi\Omega^D\Psi^T)\end{array}
	 \label{eq:ineq_JL}
	\end{equation}
Note that \eqref{eq:ineq_JL} is similar to \eqref{eq:ineq_J}, where $\Omega^D$ now replaces $\Omega^D$ and $\Sigma^D_{t|t}$ now replaces $\Sigma_{t|t}$. Furthermore, from the definition of $J^*(t)$ we also have \begin{align}J^*(t)&\geq\|\bar{x}_{t|t}\|^2_Q+\|\bar{u}_{t|t}\|^2_R\nonumber\\
&+\mathrm{tr}\{\begin{bmatrix}Q_L&0\\0&Q+K_{t|t}^TRK_{t|t}
\end{bmatrix}\Sigma_{t|t}^D\}\label{eq:ineq_JgeqL}
	\end{align}
Now, $\Omega_F$ should be denoted as  $\Omega_F=\{(\bar{x},\Sigma):\bar{x}\in\bar{\mathbb{X}}_F,\Sigma\preceq\bar{\Sigma}^D\}$. Assuming that $(\bar{x}_{t|t},\Sigma^D_{t|t})\in\Omega_F$, similarly to Section \eqref{app:proof_Theorem}, we derive
	\begin{equation}J^*(t)\leq
	\|\bar{x}_{t|t}\|^2_S+\mathrm{tr}\{S_T \Sigma^D_{t|t}\}+N\,\mathrm{tr}\{S_T \Psi\Omega^D\Psi^T\}\label{eq:ineq_JleqL}
	\end{equation}
Similarly to Section \ref{app:proof_Theorem}, from \eqref{eq:ineq_J}, \eqref{eq:ineq_Jgeq} and \eqref{eq:ineq_Jleq} we derive robust stability-related results. Namely, we denote $\omega=\mathrm{tr}\{S_T\Psi\Omega^D\Psi^T\}$. In view of this, we can reformulate \eqref{eq:ineq_JL}, \eqref{eq:ineq_JgeqL} and \eqref{eq:ineq_JleqL} as \eqref{eq:ineqs_J_b}, where now $\Sigma^D_{t|t}$ replaces $\Sigma_{t|t}$. The proof now follows exactly as in Section \ref{app:proof_Theorem}, proving that $\mathrm{dist}(\alpha V(\bar{x}_{t|t},\Sigma^D_{t|t}), [0,b\omega])\rightarrow 0$ as $t\rightarrow +\infty$ under condition \eqref{eq:cons_conds_on_WL}.
\section{The S-MPCc problem: proof of Corollary \ref{cor:const_gains}}
\label{app:proof_const_gains}
Concerning the proof of feasibility, it can be carried out exactly as in Section~\ref{app:proof_Theorem}, even in case the control and estimator gains $K_{h}$ and $L_h$, respectively, are not free variables, but are set to values $K_{h}=\bar{K}$ and $L_h=\bar{L}$ for all $h\geq 0$.\\
Concerning the proof of convergence, it follows equivalently to Section~\ref{app:proof_Theorem}. Again, the only difference now lies in the fact that $K_{t+k|t}=\bar{K}$ and $L_{t+k|t}=\bar{L}$ for all $t\geq 0$ and $k\geq 0$.

\bibliographystyle{plain}
\bibliography{stochastic}
	%
	%
	% % Each appendix must have a short title.
\end{document}